\documentclass[
aip,
amsmath,amssymb,
reprint,twocolumn % Comment onecolumn to override one column format %%%%%%%
]{revtex4-2}
\usepackage{graphicx}
\usepackage{graphicx}% Include figure files
\usepackage{dcolumn}% Align table columns on decimal point
\usepackage{bm}% bold math
\usepackage[T1]{fontenc}
\usepackage[usenames]{color}
\usepackage{soul}
\usepackage{color}

\usepackage{cancel}
\usepackage{amssymb}
\usepackage{amsmath}
\usepackage{xcolor}
\large
\usepackage{xcolor}
\usepackage[caption=false]{subfig}

\begin{document}
\def\bw{\begin{widetext}}
\def\ew{\end{widetext}}
\def\dg{\dagger}
\def\ddg{\ddagger}

\preprint{APS/123-QED}
\title{Coherent migration of the single excitation injected into finite-length segment of the biomolecular chain}

\author{Dalibor Chevizovich}
\email{cevizd@vin.bg.ac.rs}
\affiliation{Vinca Institute of Nuclear Sciences--National Institute of the Republic of Serbia, University of Belgrade, P.O. BOX 522, 11001, Belgrade, Serbia}

\author{Vasilije Matic}
\affiliation{Vinca Institute of Nuclear Sciences--National Institute of the Republic of Serbia, University of Belgrade, P.O. BOX 522, 11001, Belgrade, Serbia}

\author{Zeljko Przulj}
\affiliation{Vinca Institute of Nuclear Sciences--National Institute of the Republic of Serbia, University of Belgrade, P.O. BOX 522, 11001, Belgrade, Serbia}

\author{Slobodanka Galovic}
\affiliation{Vinca Institute of Nuclear Sciences--National Institute of the Republic of Serbia, University of Belgrade, P.O. BOX 522, 11001, Belgrade, Serbia}

\date{\today}% It is always \today, today,
             %  but any date may be explicitly specified

%\begin{frontmatter}

%% Title, authors and addresses

%% use the tnoteref command within \title for footnotes;
%% use the tnotetext command for the associated footnote;
%% use the fnref command within \author or \address for footnotes;
%% use the fntext command for the associated footnote;
%% use the corref command within \author for corresponding author footnotes;
%% use the cortext command for the associated footnote;
%% use the ead command for the email address,
%% and the form \ead[url] for the home page:dvk@df.uns.ac.rs
%%
%% \title{Title\tnoteref{label1}}
%% \tnotetext[label1]{}
%% \ead[url]{home page}
%% \fntext[label2]{}
%%%%\cortext[cor1]{Corresponding author}
%% \address{Address\fnref{label3}}
%% \fntext[label3]{}

\begin{abstract}
We study the migration of a single excitation excited at a structural element of a finite molecular segment, which is a part of a long biomolecular chain. The excitation cannot leave the segment and is locally coupled to thermal vibrations of the lattice, forming a self-trapped state corresponding to a nonadiabatic polaron. The time-dependent probability distribution of finding the excitation at the nodes of the segment is calculated, with particular emphasis on the role of the initial excitation position. A formal analogy is observed between the present model and continuous-time quantum walk models on finite chains with reflecting boundaries. The results reveal an asymmetry in the probability distribution for nodes symmetrically positioned with respect to the initially excited site, which arises solely from the asymmetric placement of the initial excitation within the finite segment. The only exception occurs when the initially excited node is located at the center of the segment, where the probability distribution becomes symmetric. The complex interference pattern and the absence of well-defined revivals stem from the non-equidistant spectrum of mode frequencies, leading to progressive dephasing of the constituent modes. As a result, the initially well-localized probability maximum fragments into one dominant maximum accompanied by several secondary maxima of lower intensity. These findings highlight the importance of boundary conditions and initial-state geometry in controlling quantum transport in finite molecular systems.
\end{abstract}

%%
%% Start line numbering here if you want
%%
% \linenumbers
%% main text

%%\pacs{63.20.kk, 87.15.-v, 87.15.K-}

\maketitle
%%%%%%%%%%%%%%%%%%%%%%%%%%%%%%%%%%%%%%%%%%%%%%%%%%%%%%%%%%%%%%%%%%

%%%%%%%%%%%%%%%%%%%%%%%%%%%%%%%%%%%%%%%%%%%%%%%%%%%%%%%%%%%%%%%%%%

\section{Introduction}

Biomolecules participate in numerous physiological processes taking place within living cells. For example, biomolecules often serve as "bridges" through which charge or energy quanta are transferred, as in the process of photosynthesis. Other examples include processes in which a molecule functions as a carrier of information conveyed to another molecule, such as bioinformation encoding, storage, and transfer, as well as molecular recognition processes. Biomolecules may also participate in processes in which the molecule, or one of its parts, acts as a trigger or regulator of other biochemical reactions, as in the case of enzymes and catalysts \cite{Voet, Lehninger, Dauxois, Frohlich, LambertNP2013, ChenACIE, CruzeiroLTP, AlvarezFQST2024}. The outcome of such processes depends not only on the intrinsic properties of the biomolecule, such as its composition or geometry, but also on the physical state of its local segments at the moment when one molecule encounters another.

Various mechanisms may affect the functioning of biomolecular chains (BmC). In particular, the functionality of a molecular chain (MC) may be impaired if the chain undergoes significant structural damage, for instance, chain breakage caused by high-energy ionizing radiation. However, structural damage is not the only mechanism capable of disrupting biomolecular functionality. For example, the injection (or emergence) of an excitation into a particular structural element (SE) of a BmC may locally alter the physical properties of the chain, such as the local distribution of charge or electric dipole moments. If the excitation persists on the structural element with sufficiently high probability for a sufficiently long time, the functionality of the active center containing this SE may be impaired or even completely disrupted. The active center of a BmC is usually not limited to a single SE. More commonly, it is a finite-length segment of the MC (molecular segment, MS), responsible for a particular biochemical process involving the MC \cite{Voet, Lehninger}.

Furthermore, resonant interactions between neighboring SEs, combined with quantum coherence effects, may cause the excitation to delocalize and subsequently reappear at a distant site \cite{ LambertNP2013, CruzeiroLTP, AlvarezFQST2024, CDAIPAdv2026}. Consequently, changes in the local physical properties of the BmC may occur not only at the site where the excitation was initially induced, but also at locations far from it. Therefore, the excitation does not necessarily have to be generated directly at the active center of the BmC. Instead, it may migrate toward the active center, thereby affecting the biochemical functionality of the MC. 

In realistic biological environments, BmCs participate in biochemical processes at temperatures at which their SEs undergo thermal oscillations around their equilibrium positions. Under such conditions, the interaction between the induced excitation and thermal oscillations of the SEs cannot be neglected and, in general, becomes a source of quantum decoherence, thereby suppressing resonant transport over longer intramolecular distances.

Interestingly, under certain conditions, some types of excitations in BmCs may form stable self-trapped (ST) quantum states capable of maintaining their stability despite interactions with thermal oscillations of the SEs of the molecular chain. In such cases, the interaction between the excitation and thermal oscillations leads to the formation of a stable ST state. During this process, the phonons of the MC become renormalized, while the excitation becomes "dressed" by a cloud of virtual phonons and acquires properties significantly different from those of a bare excitation \cite{ZdCeND, HolsteinAP1, HolsteinAP2, LF}. If the residual interaction between the ST excitation and the renormalized phonons is sufficiently weak, the excitation may migrate through the structure as an almost free particle while preserving quantum coherence \cite{ZdCeND, AK, PouthierJCP132}. Experimental \cite{PouthierPRL, CareriPRL51, BlanchetPRL54}, numerical \cite{HammTsironisEPJST147, HammTsironis}, and theoretical studies \cite{ZdCeND, AK, PouthierJCP132, PouthierPRE2008} suggest that such states may arise for vibron excitations in the amide-I region of polypeptide chains, such as crystalline acetanilide and proteins with $\alpha$-helix secondary structure. Similar processes may occur for electronic excitations in proteins, DNA, and RNA, although the nature of the ST electronic state differs from that of the ST vibron \cite{ChenACIE, CruzeiroLTP, CastroNetoCaldeira}. The properties of the ST vibron correspond to those of a small (nonadiabatic) polaron, whereas the ST electronic excitation exhibits properties characteristic of an adiabatic polaron. Due to their ability to preserve coherence during migration, ST excitations are regarded as potential carriers of energy, charge, or information along biomolecular chains. However, relatively little attention has been paid to the possible disruption of biomolecular functionality caused by the induction of an excitation on a structural element and its resonant transfer to a functional center (active site) of the chain \cite{CDAIPAdv2026}. This highlights the importance of studying excitation migration along BmCs, including the characteristic transfer times to distant sites and the residence times of excitations at particular SEs.

In Ref.~\cite{CDAIPAdv2026}, the dynamics of an excitation injected into a finite-length segment of a long homogeneous MC was investigated. It was assumed that, due to its interaction with thermal oscillations of the SEs of the MC, the excitation forms a nonadiabatic polaron. Particular attention was devoted to the probability of finding the excitation at sites located far from the initially excited site, with the aim of assessing the possibility of long-distance excitation migration. The influence of the environmental temperature, with which the MC is in thermodynamic equilibrium, as well as several structural parameters characteristic of polypeptide BmCs, on this probability and on the excitation residence time was also examined.

Another important aspect of excitation migration concerns the manner in which the excitation propagates through the molecular segment. While Ref.~\cite{CDAIPAdv2026} analyzed the probability of finding the excitation at distant sites, the spatiotemporal evolution of the excitation probability distribution throughout the entire molecular segment has not yet been investigated. In particular, it is important to understand how the migration process depends on the position of the initially excited site, whether the probability distribution remains sufficiently localized to preserve a well-defined propagation front or gradually develops a more complex interference pattern, and whether the finite geometry of the molecular segment may induce an asymmetric probability distribution despite the local symmetry of the homogeneous molecular chain.

In the present work, the primary objective is to investigate how the migration of an excitation through a finite molecular segment depends on the position of the initially excited site. To this end, we analyze the coherent spatiotemporal evolution of the excitation probability distribution over all sites of the segment, with particular attention paid to whether this probability distribution remains localized or gradually spreads over the molecular segment, indicating excitation delocalization over an increasing number of sites. The analytical formalism developed in Ref.~\cite{CDAIPAdv2026}, which provides the excitation probability at an arbitrary site of the molecular segment, enables such a comprehensive analysis and provides deeper insight into the physical mechanisms governing coherent excitation transport in finite molecular systems. As in Ref.~\cite{CDAIPAdv2026}, the excitation is assumed to form a nonadiabatic polaron due to its interaction with thermal oscillations of the structural elements of the molecular chain, while the residual interaction between the polaron and the renormalized phonons is neglected.

\section{Theoretical model}

\begin{figure}[h]
	\begin{center}
		\includegraphics[width=80mm]{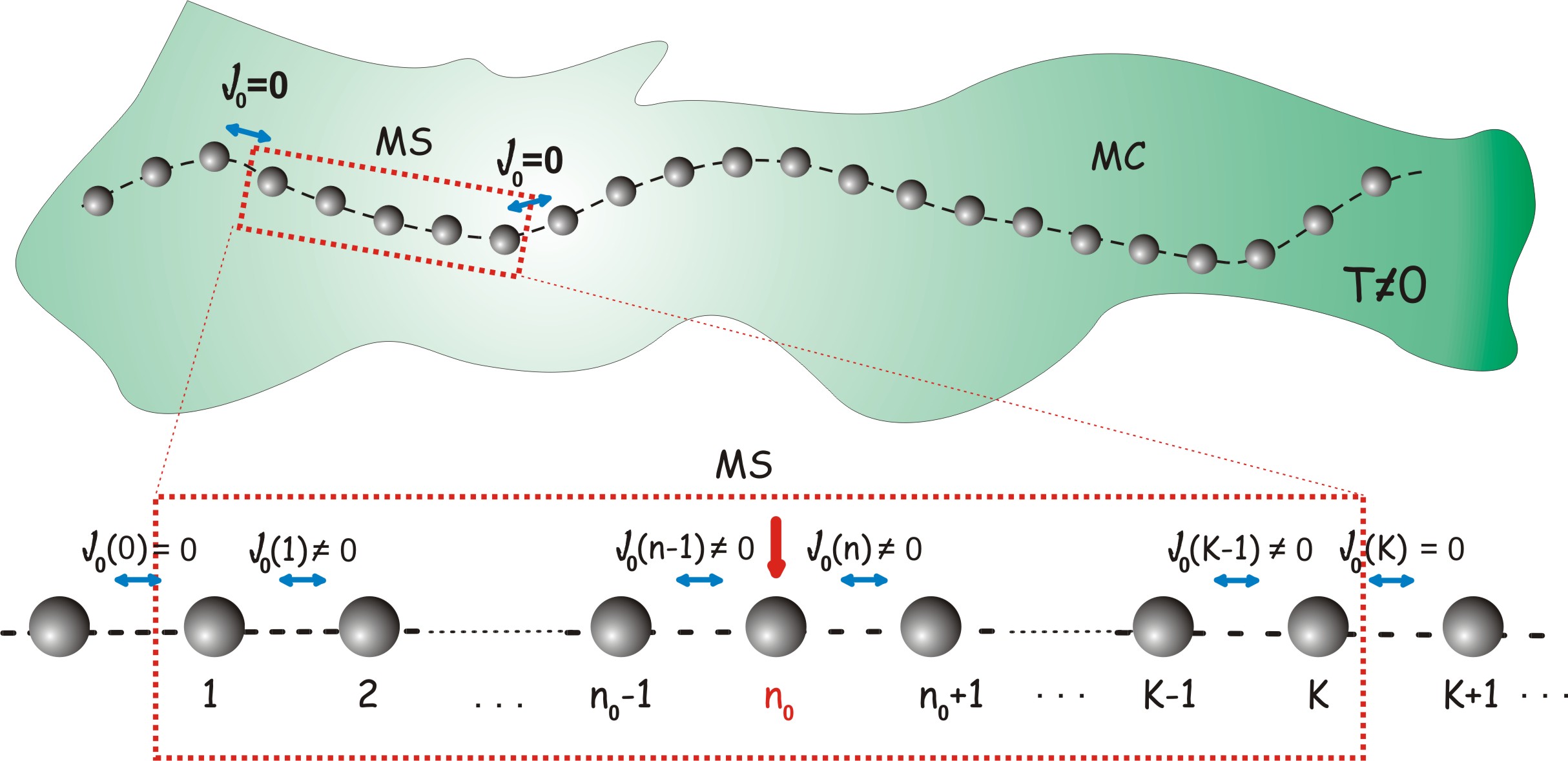}
		\caption{Schematic representation of the considered molecular structure. The excitation is initially induced at the SE labeled by $n_0$. The initially excited SE belongs to the MS, which forms part of a long MC.}\label{fig01}
	\end{center}
\end{figure}

We consider a molecular chain composed of identical structural elements regularly arranged along the chain. The SEs interact through "elastic forces", forming an infinite one-dimensional lattice. The MC is assumed to be in thermal equilibrium with its environment (thermal bath). A molecular segment, consisting of a finite number of SEs, is embedded within the MC. At the initial moment, $t_0=0$, an excitation is induced at a particular SE of the MS, as illustrated by the rectangular region in the lower part of Fig.~\ref{fig01}.

%============================= Fig. 1 =================================
%============================= Fig. 1 =================================

Since the entire chain undergoes thermal oscillations, the excitation effectively interacts with the phonon spectrum of an infinite MC. On the other hand, due to resonant interactions between neighboring SEs, the excitation may delocalize over the entire structure and, at later times, appear at any node. Furthermore, we assume that an excitation induced at a given SE of the MS can migrate along the segment, but cannot leave it \cite{CDAIPAdv2026}. This constraint is implemented in the model by setting the transfer--integrals between the boundary SEs of the MS and the adjacent SEs belonging to the remainder of the chain equal to zero (Fig.~\ref{fig01}). The rest of the chain affects the excitation only indirectly through coupling to the phonon subsystem.

To describe excitation self-trapping in such a structure, we employ the Holstein model of a molecular crystal \cite{ZdCeND, HolsteinAP1, HolsteinAP2, DevreeseRPP2009, KalosakasPRB}, adapted to the case of a finite-size MS \cite{CDAIPAdv2026}. The corresponding Hamiltonian reads:

\begin{equation}\label{PocHam}
	\hat{H}=\hat{H}_{\text{exc}}+\hat{H}_{\text{ph}}+\hat{H}_{\text{exc--ph}}
\end{equation}

\noindent where

\begin{equation}\label{PocHamExc}
	\hat{H}_{\text{exc}}=\mathcal{E}_0\sum_n\hat{a}^{\dag}_n\hat{a}_n-\sum_n\hat{a}^{\dag}_n\bigg(J_0(n)\hat{a}_{n+1}+J_0(n-1)\hat{a}_{n-1}\bigg)
\end{equation}

\begin{equation}\label{PocHamPh}
	\hat{H}_{\text{ph}}=\sum_q\hbar\omega_q\hat{b}^{\dag}_q\hat{b}_q
\end{equation}

\begin{equation}\label{PocHamExcPh}
	\hat{H}_{\text{exc--ph}}=\tfrac{1}{\sqrt{N}}\sum_n\sum_qF_q\mathrm{e}^{iqnR_0}\hat{a}^{\dag}_n\hat{a}_n(\hat{b}_q+\hat{b}^{\dag}_{-q})
\end{equation}

\noindent Here, $\hat{a}^{\dg}_n$ ($\hat{a}_n$) are the creation (annihilation) operators of the excitation on the $n$--th SE of the MC (the index $n$ enumerates all structural elements along the entire chain), while $\hat{b}^{\dg}_q$ ($\hat{b}_q$) are the creation (annihilation) operators of phonons with wave number $q$. The energy required to excite the corresponding excitation mode on a particular SE of the MC is denoted by $\mathcal{E}_0$. 

According to incoherent neutron scattering studies of ACN \cite{Barthes1989} or IR pump--probe experiments \cite{Edler2002}, vibron self--trapping in this system arises exclusively from coupling with optical phonons. For this reason, we restricted our study to the interaction of a single excitation with optical phonons only. Consequently, the excitation--phonon coupling constant is $F_q=F=\chi\sqrt{\frac{\hbar}{2M\omega_0}}$. The parameter $\chi$ is the excitation--phonon coupling parameter, which characterizes the strength of the interaction between the excitation and phonon modes \cite{ZdCeND, LF, AK,PouthierJCP132, PouthierPRL, PouthierPRE2008, KalosakasPRB, YarkonyJCP, HennigPRB}. The characteristic phonon frequency is $\omega_q=\omega_0=2\sqrt{\frac{\kappa}{M_0}}$. Here, $\kappa$ is the MC stiffness constant and $M_0$ is the mass of the SE.

The transfer--integral between the $n$-th SE and the $(n+1)$-th SE of the MC is denoted by $J_0(n)$ (Fig.~\ref{fig01}). In the case of vibron excitations, this quantity corresponds to the energy of resonant dipole--dipole interactions between neighboring SEs. Its dependence on the node index $n$ in Eq.~(\ref{PocHamExc}) is purely formal. In our model, $J_0(n)=\text{const.}=J_0$ for any pair of neighboring SEs within the segment. For the transfer--integrals that govern hopping between the edge SEs of the segment and their nearest neighbors in the rest of the chain, we set $J_0(n)=0$ (see Fig.~\ref{fig01}). At the same time, the summation over $n$ in Eqs.~(\ref{PocHamExc}) and (\ref{PocHamExcPh}) can formally be taken over the entire molecular chain.

To transform the system into the picture of an ST excitation (or, more generally, into the polaron representation), we apply the Lang--Firsov unitary transformation \cite{ZdCeND, LF, DevreeseRPP2009, YarkonyJCP, Rashba, Mahan} (LFuT): $$\hat{U}_{\text{LF}}=\mathrm{e}^{-\sum_n\hat{a}^{\dag}_n\hat{a}_n\hat{S}_n},\; \hat{S}_n=\tfrac{1}{\sqrt{N}}\sum_q \tfrac{F^*_q}{\hbar\omega_q} \mathrm{e}^{-iqnR_0}\left(\hat{b}_{-q}-\hat{b}^{\dag}_q\right)$$ The transformed Hamiltonian $\hat{H}_{\text{LF}}=\hat{U}_{\text{LF}}\hat{H}\hat{U}^{\dag}_{\text{LF}}$ takes the form

\begin{align}\label{HLF}
	\hat{H}_{\text{LF}}&=\mathcal{E}_R\sum_n\hat{a}^{\dag}_n\hat{a}_n-\sum_nJ_0(n)\hat{a}^{\dag}_n\hat{a}_{n+1}\hat{T}_+(n)-\nonumber\\
	&-\sum_nJ_0(n-1)\hat{a}^{\dag}_n\hat{a}_{n-1}\hat{T}_-(n)+\sum_q\hbar\omega_q\hat{b}^{\dag}_q\hat{b}_q+\hat{O}_{\text{rest}}\nonumber
\end{align}

\noindent Here, $\hat{a}^{\dg}_n$ ($\hat{a}_n$) are the creation (annihilation) operators of the ST excitation on the $n$--th SE, and $\hat{b}^{\dg}_q$ ($\hat{b}_q$) are the creation (annihilation) operators of the renormalized phonons. The energy of the ST excitation is $\mathcal{E}_R=\mathcal{E}_0-\mathcal{E}_b$, where the excitation binding energy (energy shift) is given by $\mathcal{E}_b=-\frac{1}{N}\sum_q\frac{|F_q|^2}{\hbar\omega_q}$ \cite{ZdCeND, LF, YarkonyJCP}. The dressing mechanism is incorporated via the application of the LFuT, which leads to the appearance of the nonlinear operators $\hat{T}_{\pm}(n)=\mathrm{e}^{\hat{S}_{n\pm 1}-\hat{S}_n}$. The LFuT provides an exact diagonalization of the Hamiltonian in the limit $J_0\to 0$. However, when $J_0 \neq 0$, a nonlinear coupling  $\hat{a}^{\dag}_n\hat{a}_{n\pm 1}\hat{T}_{\pm}(n)$ remains, which may give rise to dissipation processes and incoherent motion of the ST excitation. The term $\hat{O}_{\text{rest}}$ represents phonon--mediated "residual" interaction between two ST excitations, which can be neglected in the single--excitation case.

To account for the influence of thermal fluctuations on the properties of the ST excitation, we apply the mean-field procedure and introduce the corresponding mean-field Hamiltonian of the system \cite{ZdCeND, LF, YarkonyJCP}. In particular, we define the effective mean-field Hamiltonian $\hat{\mathcal{H}}_0$ as $$\hat{H}_{\text{LF}}=\hat{\mathcal{H}}_0+\hat{\mathcal{H}}_{\text{rest}},$$ where $\hat{\mathcal{H}}_0=\hat{\mathcal{H}}_R+\hat{\mathcal{H}}_{\text{ph}}$,
$\hat{\mathcal{H}}_R=\left\langle \hat{H}_{\text{LF}}-\hat{\mathcal{H}}_{\text{ph}}\right\rangle_{\text{ph}}$, and
$\hat{\mathcal{H}}_{\text{rest}}=\hat{H}_{\text{LF}}-\hat{\mathcal{H}}_{\text{ph}}-\left\langle \hat{H}_{\text{LF}}-\hat{\mathcal{H}}_{\text{ph}}\right\rangle_{\text{ph}}$. The symbol $\left\langle...\right\rangle_{\text{ph}}$ denotes averaging over the ensemble of renormalized phonons, assumed to be in thermal equilibrium with the thermal bath. Finally, the Hamiltonian of the ST excitation is:

\begin{equation}\label{HSP}
	\hat{\mathcal{H}}_R=\mathcal{E}_R\sum_n\hat{a}^{\dag}_n\hat{a}_n-J_0\mathrm{e}^{-W(T)}\sum_n\hat{a}^{\dag}_n\left(\hat{a}_{n+1}+\hat{a}_{n-1}\right)
\end{equation}

\noindent The thermal average of the nonlinear operators $\hat{T}_{\pm}(n)$ is given by $\left\langle\hat{T}_{\pm}(n)\right\rangle_{\text{ph}} =\mathrm{e}^{-W(T)}$, where the narrowing factor of $J_0$ is

\begin{equation}\label{WT}
W(T)=\tfrac{1}{N}\sum_q\frac{F^2}{(\hbar\omega_0)^2}(2n_q+1)(1-\cos(qR_0))
\end{equation}

\noindent and $n_q=\frac{1}{\mathrm{e}^{\frac{\hbar\omega_0}{k_BT}}-1}$. Although the value of transfer--integral is reduced, the fact that it remains nonzero within the segment allows the excitation to migrate along it and appear on any of its structural elements.

The term $\hat{\mathcal{H}}_{\text{rest}}$ includes residual interaction between the ST excitation and the phonon subsystem, as well as the phonon mediated polaron-polaron interaction:

\begin{align}\label{Hrest}
	\hat{\mathcal{H}}_{\text{rest}}&=\sum_nJ_0(n)\hat{a}^{\dag}_n\hat{a}_{n+1}\big(\hat{T}_+(n)-\left\langle\hat{T}_+(n)\right\rangle_{ph}\big)+\nonumber\\
	&+\sum_nJ_0(n-1)\hat{a}^{\dag}_n\hat{a}_{n-1}\big(\hat{T}_-(n)-\left\langle\hat{T}_-(n)\right\rangle_{ph}\big)+\nonumber\\
	&+\big(\hat{O}_{\text{rest}}-\left\langle\hat{O}_{\text{rest}}\right\rangle_{ph}\big)\nonumber
\end{align}

The first two terms in $\hat{\mathcal{H}}_{\text{rest}}$ describe fluctuations of the nonlinear polaron--renormalized phonon coupling operators around their thermal average values. These fluctuations are assumed to affect the polaron dynamics only weakly and are therefore neglected within the framework of the mean-field approximation. Consequently, dissipative processes associated with incoherent polaron--phonon scattering are not captured by the present model.

The Hamiltonian $\hat{\mathcal{H}}_R$, defined by Eq.~(\ref{HSP}), governs the excitation dynamics in our model and provides the mathematical basis for the subsequent analysis. To analyze the appearance of the excitation on the SEs of the MS, we follow the procedure presented in Ref.~\cite{CDAIPAdv2026}. Let us assume that, at the initial moment $t=0$, the excitation is induced on a particular SE labeled by the index $n_0$ (Fig.~\ref{fig01}). The corresponding initial state is $\left|\psi_i(0)\right\rangle= \hat{a}_{n_0}^{\dg}\left|0\right\rangle$, where $\left|0\right\rangle$  denotes the excitation vacuum state. In the absence of external influences, the state evolves in time according to $\left|\psi_i(t)\right\rangle= \mathrm{e}^{-\frac{i}{\hbar}t\hat{\mathcal{H}}_R} \hat{a}_{n_0}^{\dg}\left|0\right\rangle$. We are interested in the probability of finding the excitation at another site $n\neq n_0$ at a later time $t>0$, i.e., in the probability that the system reaches the state $\left|\psi_f(t)\right\rangle=\hat{a}^{\dg}_n\left|0(t)\right\rangle$. Here, $\left|0(t)\right\rangle=\mathrm{e}^{-\frac{i}{\hbar}t\hat{\mathcal{H}}_R}\left|0\right\rangle$ describes the time evolution of the excitation vacuum state. The corresponding probability is $p_n(t)=\left|\left\langle\psi_f(t)\middle|\psi_i(t)\right\rangle\right|^2$, where the corresponding correlation function (CF) is given by $V_n(t)=\left\langle\psi_f(t)\middle|\psi_i(t)\right\rangle=\left\langle 0\right|\mathrm{e}^{\frac{i}{\hbar}t\hat{\mathcal{H}}_R}\hat{a}_n\mathrm{e}^{-\frac{i}{\hbar}t\hat{\mathcal{H}}_R}\hat{a}_{n_0}^{\dg}\left|0\right\rangle$. In the Heisenberg picture it is given by:

\begin{equation}\label{CFVH}
	V_n(t)=\left\langle 0\right|\hat{a}_n(t)\hat{a}_{n_0}^{\dg}\left|0\right\rangle;\quad \hat{a}_{n_0}^{\dg}=\hat{a}_{n_0}^{\dg}(t=0)
\end{equation}

The CF defined by Eq.~(\ref{CFVH}) determines the probability amplitude for the transition of the excitation from the site where it was initially induced (the "zeroth" site) to a distant $n$--th site during the time interval between $t_0=0$ and a later time $t$. Mathematically, it connects the creation of the excitation at the initial site at $t=0$ with its appearance at site $n$ at time $t$. For brevity, we will hereafter refer to the CF defined by Eq.~(\ref{CFVH}) simply as the correlation function associated with site $n$ at time $t$. The subsequent analysis is based on solving the system of differential equations for the CFs $V_n(t)$ corresponding to all sites of the MC. Furthermore, we introduce the dimensionless reduced time variable $\tau=\omega_0t$. Under this transformation, time derivatives are rewritten as $\frac{d f(t)}{dt}=\omega_0\frac{d f(\tau)}{d\tau}$.

In the polaron representation, it is convenient to introduce two dimensionless parameters: the coupling constant $S=\frac{\mathcal{E}_b}{\hbar\omega_0}$ which characterizes the strength of the polaron--phonon interaction, and the adiabatic parameter $B=\frac{2J_0}{\hbar\omega_0}$  determining the character of the lattice deformation involved in the self--trapping process. In the case of the vibron excitation in polypeptide BmCs, one typically has $B\ll 1$, and $S/B\gg 1,$ which corresponds to the regime of a small (non--adiabatic) polaron. These conditions are usually realized in structures with narrow excitation bands or for sufficiently small values of the transfer--integral $J_0$. In this regime, the excitation and the lattice deformation become strongly coupled and form a new quasiparticle, namely a dressed excitation characterized by a renormalized effective mass and a narrowed energy band. The associated lattice distortion is localized over only a few lattice sites. By adopting $\mathcal{E}_0=0$, the system of differential equations for the CFs takes the form:

\begin{align}\label{dV}
	i\frac{d{V}_1(\tau)}{d\tau}&=-SV_1(\tau)-\tfrac{B}{2}\mathrm{e}^{-W(\theta)}V_2(\tau)\nonumber\\
	i\frac{d{V}_2(\tau)}{d\tau}&=-SV_2(\tau)-\tfrac{B}{2}\mathrm{e}^{-W(\theta)}(V_1(\tau)+V_3(\tau))\nonumber\\
	..............&............................................\nonumber\\
	i\frac{d{V}_n(\tau)}{d\tau}&=-SV_n(\tau)-\tfrac{B}{2}\mathrm{e}^{-W(\theta)}(V_{n-1}(\tau)+V_{n+1}(\tau))\nonumber\\
	..............&............................................\\
	i\frac{d{V}_{K-1}(\tau)}{d\tau}&=-SV_{K-1}(\tau)-\tfrac{B}{2}\mathrm{e}^{-W(\theta)}(V_{K-2}(\tau)+V_K(\tau))\nonumber\\
	i\frac{d{V}_K(\tau)}{d\tau}&=-SV_K(\tau)-\tfrac{B}{2}\mathrm{e}^{-W(\theta)}V_{K-1}(\tau)\nonumber
\end{align}

\noindent Here, $\theta=\frac{k_BT}{\hbar\omega_0}$ is temperature, normalized on the characteristic phonon energy $\hbar\omega_0$ and $W(\theta)=S\mathrm{coth}\left(1/2\theta\right)$. At the initial moment, the excitation is induced at the $n_0$ node:

\begin{equation}\label{IntCond}
	|V_{n_0}(\tau=0)|^2=1,\quad |V_{n\neq n_0}(\tau=0)|^2=0
\end{equation}

Two equivalent ways of indexing the SEs within the MS are used in the present work. In the first (global) scheme, the SEs are numbered consecutively from the left edge of the MS to the right edge, as shown in Fig.~\ref{fig01}. The leftmost SE is labeled by $1$, while the rightmost SE is labeled by $K$, so that the site index takes values $n\in\{1,\dots,K\}$. The initially excited node is labeled with $n_0$. This indexing is used to write the system of equations Eqs.~(\ref{dV}). it is most convenient for graphical representation of the results and for possible numerical analysis of the system.

\begin{figure}[h]
	\begin{center}
		\includegraphics[width=80mm]{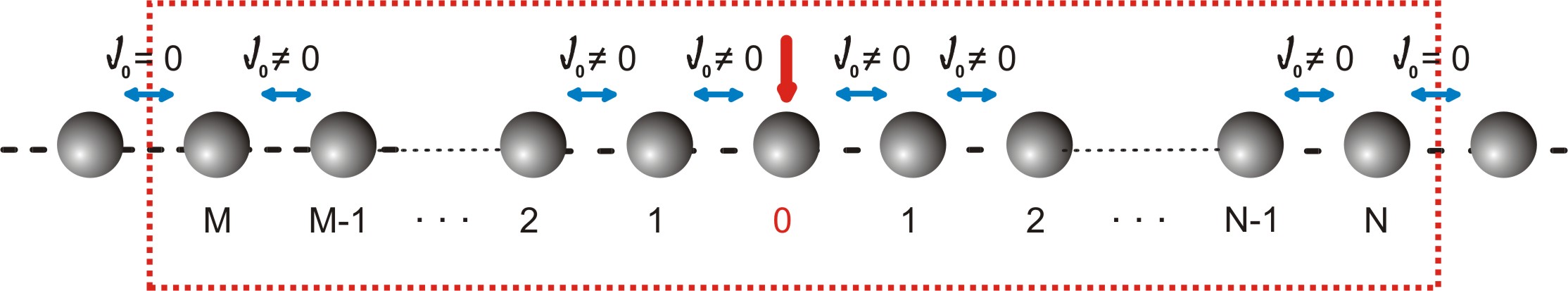}
		\caption{A schematic representation of the renumbered MS. The SE where the excitation is initially induced is labeled as $0$. On the left side of the zeroth SE, there are $M$ SEs numbered from $1$ to $M$. On the right side, there are $N$ SEs numbered from $1$ to $N$.}\label{fig02}
	\end{center}
\end{figure}

For the analytical treatment, it is convenient to introduce an alternative site indexing scheme using the initially excited site as the origin. The SE at which the excitation is induced at the initial moment $\tau_0=0$ is labeled by $n_0=0$, while all other SEs are indexed relative to this site. The index $n$ is used both for sites located to the right ($n=1,2,\dots,N$) and to the left ($n=1,2,\dots,M$) of the initially excited site, as shown in Fig.~\ref{fig02}. To distinguish between the two sides, we introduce superscripts $(R)$ and $(L)$ for the corresponding correlation functions, their Laplace transforms, and the excitation probabilities. Although this representation is less intuitive geometrically, it enables a direct connection between the Laplace-transformed CFs and the corresponding Chebyshev polynomials of the second kind. For this reason, the first indexing scheme is used in the graphical presentation of the results, while the second is employed in the analytical derivations.

%============================= Fig. 2 =================================
%============================= Fig. 2 =================================

Applying the Laplace transform $\mathcal{L}\{f(\tau)\}=\tilde{f}(s)$, where $s\in\mathbb{C}$ is the complex frequency, transforms the system of differential equations Eqs.~(\ref{dV}) into a system of algebraic equations for the $\tilde{V}_n(s)$. With the substitution $x=\frac{2}{B}\mathrm{e}^{W(\theta)}\left(-is+ S\right)$, the equations become analytically solvable. The resulting expressions for $\tilde{V}_n(x)$ are given in terms of ratios of modified Chebyshev polynomials of the second kind, $D_n(x)$ \cite{CDAIPAdv2026}. Here, $D_n(x)=U_n(2x)$, where $U_n(x)$ are the "standard" Chebyshev polynomials of the 2nd kind. Due to the adopted indexing scheme, it is convenient to distinguish between CFs on the right and left sides of the initially excited site $n_0=0$, denoted as $\tilde{V}^{(R)}_n(x)$ and $\tilde{V}^{(L)}_n(x)$, respectively. The system Eq.~(\ref{dV}) is then solved separately for the two sets of functions, following the procedure of \cite{CDAIPAdv2026}, yielding:

\begin{subequations}\label{Vnx}
	\begin{align}
		\tilde{V}^{(R)}_n(x)&=-\tfrac{2i}{B}\mathrm{e}^{W(\theta)}\frac{D_{N-n}(x)D_M(x)}{D_{M+N+1}(x)}V_0(0)\\
		\tilde{V}^{(L)}_n(x)&=-\tfrac{2i}{B}\mathrm{e}^{W(\theta)}\frac{D_{M-n}(x)D_N(x)}{D_{M+N+1}(x)}V_0(0)
	\end{align}
\end{subequations}
% subequations omogucuje numeraciju (9a) i (9b)

We note that the initial condition Eq.~(\ref{IntCond}) is incorporated into the obtained solutions Eqs.~(\ref{Vnx}) through the choice of the parameters $N$ and $M$, for a given total number of sites $K=N+M+1$. The choice of these parameters determines which site of the structure is initially excited.

\begin{figure}[h]
	\begin{center}
		\includegraphics[width=60mm]{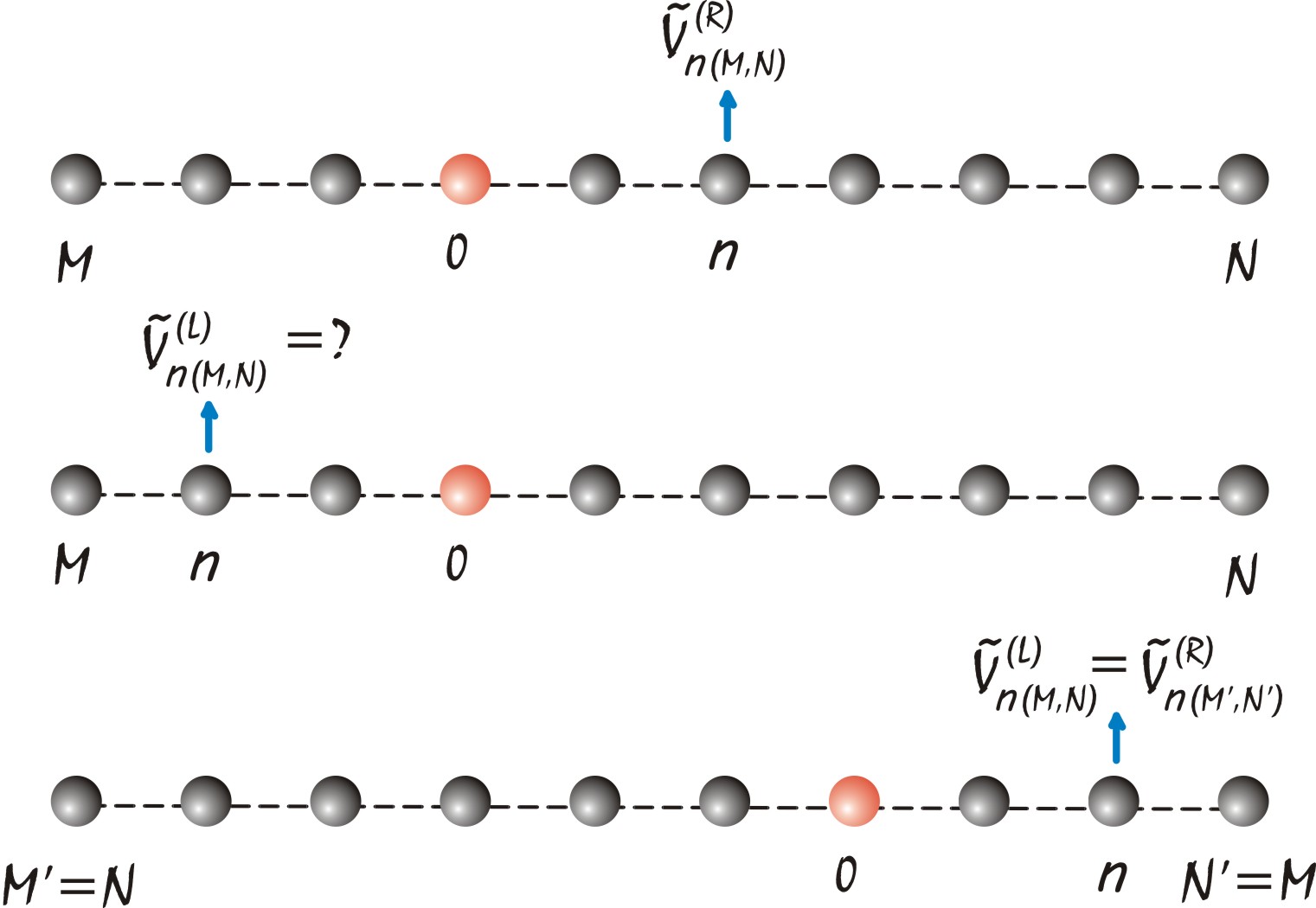}
		%\vspace{-3mm}
		\caption{A schematic illustration of the rule allowing one to obtain $V^{(L)}_n(x)$ for a $(M,N)$ structure, from $V^{(R)}_n(x)$ of the corresponding $(M'=N,\,N'=M)$ structure.}\label{fig03}
	\end{center}
	%\vspace{-5mm}
\end{figure}

It is useful to observe that it is not necessary to explicitly present both expressions for $\tilde{V}^{(R)}_n(x)$ and $\tilde{V}^{(L)}_n(x)$. Indeed, the expression for $\tilde{V}^{(L)}_n(x)$ for a structure $(M,N)$ can be directly obtained from the expression for $\tilde{V}^{(R)}_n(x)$ corresponding to the equivalent structure $(M'=N,\,N'=M)$, in which the considered $n$-th site is shifted to the right with respect to the initially excited site (as illustrated in Fig.~\ref{fig03}). This observation also holds for the correlation functions $V^{(R,L)}_n(\tau)$. In the following, we derive the expressions for $V^{(R)}_n(\tau)$, while the corresponding expressions for $V^{(L)}_n(\tau)$, which are required for the numerical illustrations of the results, will not be presented explicitly.

%============================= Fig. 3 =================================
%============================= Fig. 3 =================================

In order to determine the correlation functions $V_n(\tau)$ via the inverse Laplace transform, the ratio of polynomials in Eq.~(\ref{Vnx}) must first be decomposed into partial fractions: $$\frac{D_{N-n}(x)D_M(x)}{D_{M+N+1}(x)}=\sum_{k=1}^K\frac{C_k}{x-x_k}$$ where $x_k$ are the zeros of the $D_{M+N+1}(x)$. Such a decomposition is possible due to the properties of the Chebyshev polynomials of the 2nd kind, since each polynomial $D_n(x)$ has exactly $n$ simple zeros. Using the trigonometric representation of the modified Chebyshev polynomials 

\begin{equation}\label{DTrig}
D_n(x)=\frac{\sin((n+1)\phi)}{\sin(\phi)}
\end{equation}

\noindent where $x=2\cos(\phi)$, more general expressions for the correlation functions can be obtained. The zeros of the $D_n(x)$ can be represented as

\begin{equation}\label{xtrig}
	x_k=2\cos\bigg(\frac{k\pi}{n+1}\bigg); \quad k\in \left\{1,2,...,n\right\}
\end{equation} 

\noindent Finally, the ratio of polynomials from Eq.~(\ref{Vnx} a)) can be written in the following form:

\begin{subequations}\label{Dkolic}
\begin{align}
\frac{D_{N-n}(x)D_M(x)}{D_K(x)}&=\tfrac{2}{K+1}\sum_{k=1}^K\frac{\Psi^{\left\{M,N\right\}}_{n,0}(\phi_k)}{x-2\cos(\phi_k)}\\
\Psi^{\left\{M,N\right\}}_{n,0}(\phi_k)&=\psi^{\left\{M,N\right\}}_n(\phi_k)\psi^{ \left\{M,N\right\}}_0(\phi_k)
\end{align}
\end{subequations}

\noindent where  $\psi^{\left\{M,N\right\}}_n(\phi_k)=\sin\left[(M+n+1)\phi_k\right]$, $\phi_k=\frac{k\pi}{K+1}$, $k\in\left\{1,2,...,K\right\}$ and $K=M+N+1$. On the same way, one can transform the ratio from Eq.~(\ref{Vnx} b)). Applying the inverse Laplace transform $\mathcal{L}^{-1}\left\lbrace\tilde{f}(s)\right\rbrace=f(\tau)$ the correlation functions in the time domain attain the form:

\begin{subequations}\label{VntauN}
	\begin{align}
	V_n(\tau)&=\tfrac{2}{K+1}\mathrm{e}^{iS\tau}V_0(0)\sum_{k=1}^K\mathrm{e}^{i\Omega_k\tau}\Psi^{\left\{M,N\right\}}_{n,0}(\phi_k)\\
	\Omega_k&=B\mathrm{e}^{-S\mathrm{coth}(1/2\theta)}\cos(\phi_k)
	\end{align}
\end{subequations}

\noindent Equation~(\ref{VntauN}a) shows that each correlation function $V_n(\tau)$ is expressed as a coherent superposition of discrete modes $\mathrm{e}^{i\Omega_k\tau}$, whose characteristic frequencies $\Omega_k$ form a discrete spectrum determined by the length of the molecular segment and independent of the position of the initially excited site. On the other hand, the influence of the position of the initially induced node is expressed through the "weighting coefficients" $\Psi^{\left\{M,N\right\}}_{n,0}(\phi_k)$, which determine the relative contribution of each discrete mode to the corresponding correlation function. Unlike the discrete modes, the weighting coefficients depend on the position of the site where the probability of finding the excitation is observed $n$ relative to the initially excited site, $n_0=0$ (i.e. on its position relative to the ends of the molecular segment).

Using Eq.~(\ref{VntauN}) we can calculate {\it the probability of finding the single excitation at the $n$-th site at time $\tau$ when the excitation is initially induced at site $n_0$} (hereafter, will be referred to as the {\bf excitation probability distribution}):

\begin{equation}\label{pntau}
	p_n(\tau)=|V_n(\tau)|^2
\end{equation}

From Eqs.~(\ref{VntauN}) and (\ref{pntau}), the probability $p_n(\tau)$ contains not only the individual contributions of the discrete modes but also their mutual interference terms. Consequently, the excitation probability distribution reflects the coherent interference among all discrete modes participating in the dynamics.

% p_n(\tau) NIJE KOHERENTNA SUPERPOZICIJA modova e^{i\Omega_k\tau}, to je V_n(\tau)
% p_n(\tau) sadrzi "koherencije"-nedijagonalne clanove koji odgovaraju matricnim koeficijentima iz V_n, ali sa razlicith mjesta
% p_n(\tau) sadrzi interferencione (cross) clanove izmedju razlicitih parova modova! (npr/ i(\Omega_k1-\Omega_k2)\tau)

% $\beta_j=S+\frac{B}{2}\mathrm{e}^{-W(\theta)}x_j$, $W(\theta)=S\mathrm{coth}(1/2\theta)$

\section{Results and discussion}\label{Results}

Eqs.~(\ref{VntauN} a, b) and (\ref{pntau}) enable the analysis of the time-dependent probability of finding the excitation at a given ($n$-th) site of the MS when the excitation is initially induced at the site $n_0$. According to Eqs.~(\ref{VntauN}) and (\ref{pntau}), the quantity $p_n(\tau)$ can be interpreted as a coherent superposition of several harmonics, each associated with a particular value of $\Omega_k$. Since the time dependence of $p_n(\tau)$ for fixed $n$ exhibits a spectral-like structure, we introduce the term \textbf{temporal probability spectrum} (TPS).

The values of the physical parameters used in the numerical calculation correspond to those typical for an amide-I excitation in a polypeptide BMC, such as proteins with an $\alpha$-helical secondary structure or quasi-one-dimensional molecules like ACN. These structures are among the most experimentally and theoretically studied systems in the context of polaron-mediated intramolecular energy transfer. Available experimental and theoretical data suggest that the value of the transfer-integral typically lies within the range $J_0\approx (0.5-0.97)$ meV and that the vibron-phonon coupling parameter is $\chi\approx (32-62)$ pN. The stiffness constant is $\kappa\approx (13-20)$ N/m for ACN, and $\kappa\approx (39-58)$ N/m for $\alpha$-helical proteins. The effective masses of the SEs are approximately $M_0\approx 2.25\cdot10^{-25}$ kg for ACN and $M_0\approx 5.7\cdot 10^{-25}$ kg for the $\alpha$--helix \cite{CareriPRL51, AK, HennigPRB, Nevskaya, FalvoPouthier, HammEdlerPRB73}. Consequently, the characteristic angular frequency is $\omega_0\approx 10^{13}$ 1/s and the dimensionless parameters $S$ and $B$ typically take the values $S\approx 0.3$ and $B\approx 0.1$. It is important to note that the precise numerical values of the parameter $\chi$ are unknown. This parameter cannot be measured directly in experiments, whether the excitation is a vibron or an injected electron \cite{PouthierJCP132,KalosakasPRE}. Instead, $\chi$ is inferred indirectly from other directly measurable quantities that depend on it. Furthermore, its value is model-dependent, as it is determined based on a pre-assumed theoretical framework, which adds an additional layer of uncertainty.

%Sto nam daje slobodu da vrijednost parametra $S$ biramo iz nesto sireg intervala, umjesto da uzimamo neku tacno odredjenu vrijednost. 

In Ref.~\cite{CDAIPAdv2026}, the general properties of the TPS in homogeneous finite-length MSs were analyzed, including the influence of temperature, system parameters, and the position of the initially excited site on the resulting excitation probability distribution. Here we provide further insight into certain aspects of the "migration" of excitation through such a structure at temperatures relevant to biological processes. Accordingly, in the following calculations we restrict our analysis to the room-temperature regime, with all calculations performed for $\theta=4$, $S=0.3$ and $B=0.1$.

\subsection{Two preliminary observations\\}

Before analyzing the excitation migration in detail, we first discuss two general features that follow directly from the obtained analytical solutions. These preliminary observations provide the physical framework for understanding the results presented in the following sections.\\

\noindent {\it A) Formal similarity with continuous-time quantum walk}\\

Let us note that the expressions for the correlation functions, given by Eq.~(\ref{VntauN}a), exhibit a mathematical structure closely analogous to that encountered in continuous-time quantum walk (CTQW) models on finite chains with reflecting boundary conditions \cite{Kempe2003, MB2011}. In both cases, the dynamics is governed by a discrete set of normal modes determined by the geometry and boundary conditions of the system, while the initially excited site determines the projection of the initial state onto these modes. 
%Ovo znaci slijedece: ako u $t_0=0$ pobudimo npr $n_0=3$ cvor, nastala "oscilacija" se matematicki opisuje kao linearna kombinacija svih normalnih modova. Pri tome, neki modovi ce imati veci udio, drugi manji, a to zavisi od polozaja inicijalno pobudjenog cvora! To se zove projekcija stanja na modove. 
In the present formulation, the corresponding spectrum is determined by the zeros of the polynomial $D_K(x)$. Furthermore, the appearance of Chebyshev polynomials of the 2nd kind in the analytical expressions for the probability distribution additionally supports this analogy in light of the results reported in \cite{FussArXiv2007, AperarXiv2024}. At present, CTQW models are widely employed to describe coherent quantum transport in a broad range of physical systems, including quantum-information networks, condensed-matter structures, and biomolecular chains. In particular, they provide a minimal framework for understanding coherent excitation transfer in low-dimensional systems.\\

\noindent {\it B) Asymmetry}\\

It should also be noted that the expressions in Eq.~(\ref{Vnx}) exhibit a certain asymmetry between the Laplace transforms of the correlation functions $\tilde{V}^{(L)}_n(x)$ and $\tilde{V}^{(R)}_n(x)$ corresponding to sites symmetrically positioned with respect to the initially excited site. Namely, for a structure specified by the parameters $(M,N)$, the expressions for $\tilde{V}^{(R)}_n(x)$ and $\tilde{V}^{(L)}_n(x)$ contain the same polynomial $D_K(x)$ in the denominator, while the polynomials appearing in the numerator are different: in the first case, the numerator is given by the product $D_{N-n}(x)D_M(x)$, whereas in the second case it is given by $D_{M-n}(x)D_N(x)$. As a consequence, the corresponding correlation functions $V^{(L)}_n(\tau)$ and $V^{(R)}_n(\tau)$ differ from one another, leading to different temporal probability spectra (TPSs) on the left- and right-hand sides of the initially excited site. Although the molecular segment is homogeneous and the local transfer integrals do not favor any particular direction of excitation migration--nor is there any external field that would induce directed transport--the position of the initially excited site itself may influence the emergence of asymmetry in the excitation dynamics. 

In a broader context, this behavior bears some resemblance to asymmetry effects discussed in the context of Hamiltonian quantum ratchets and transport phenomena generated by specially prepared initial states \cite{PhysRepReiman2002, LauRSC2017, Kozyrev2023}. However, unlike driven or dissipative ratchet systems that exhibit directed transport, the present model describes a closed and undriven system. Our analytical solution in terms of modified Chebyshev polynomials of the second kind provides an exact description of this effect, complementing previous studies of Hamiltonian ratchets and related finite-size transport phenomena governed by tailored initial states \cite{PhysRepReiman2002, LauRSC2017, Kozyrev2023}.

\begin{figure}[h]
	\begin{center}
		\includegraphics[width=4.2cm]{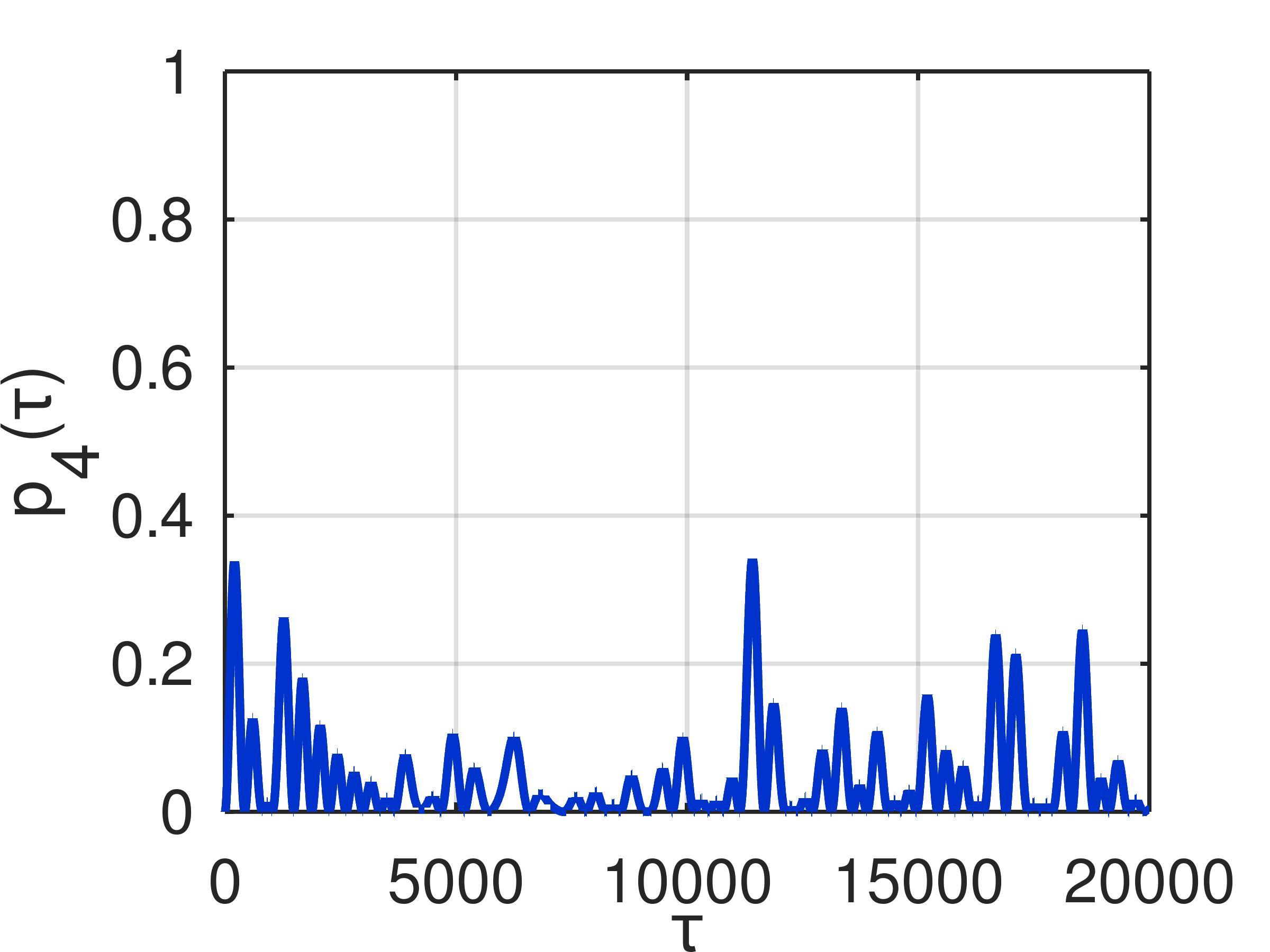}
		\includegraphics[width=4.2cm]{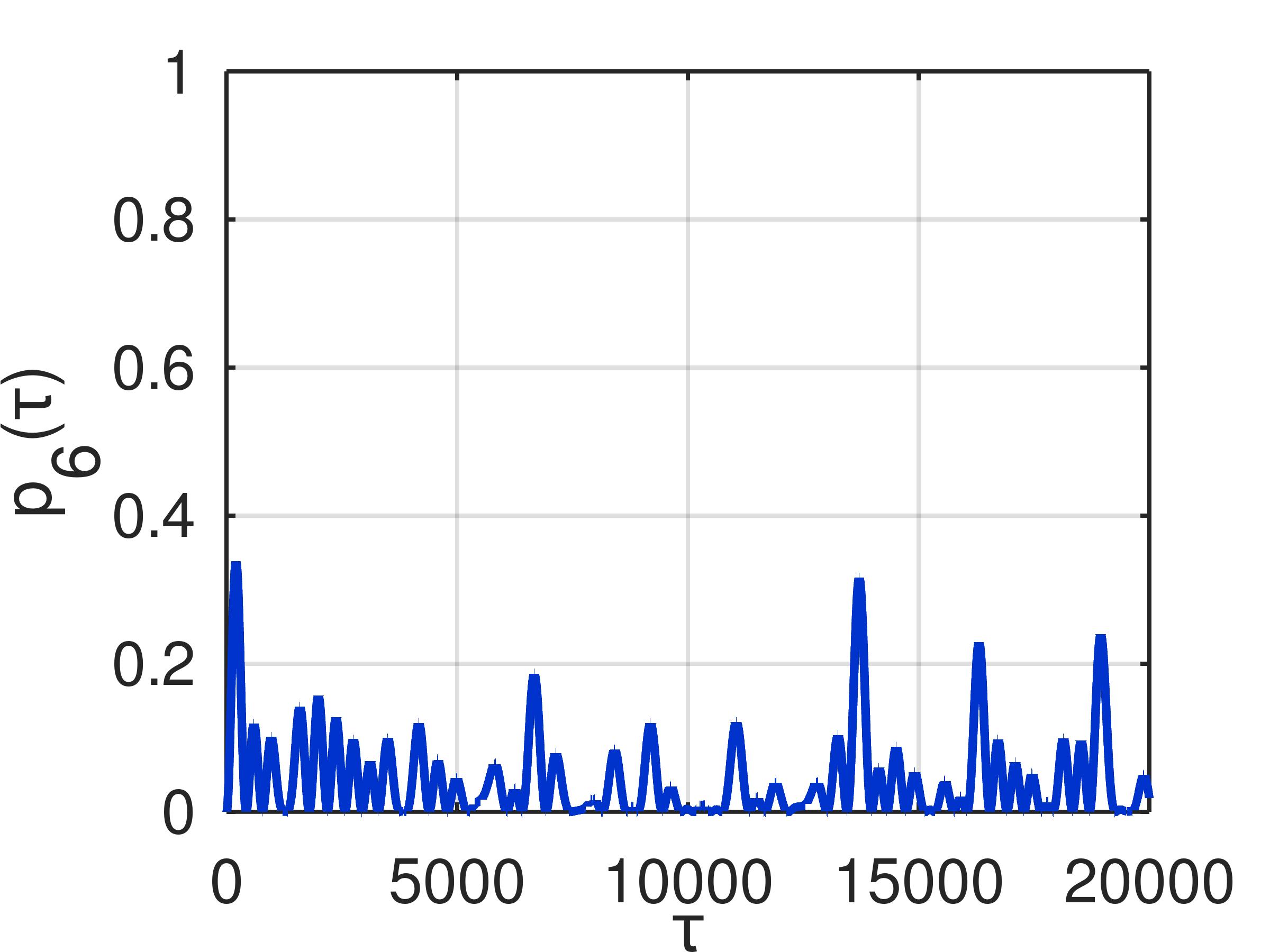}
		\caption{TPS corresponding to the two nearest-neighbor sites of the initially excited site in a MS with $M=4$, $N=16$ ($K=21$) SEs. For clarity, the site numbering in the figure follows the convention introduced in Fig.~\ref{fig01}, where the sites are numbered starting from the left edge of the MS. The initially excited site is $n_0=5$. Left panel: TPS corresponding to site $n=4$. Right panel: TPS corresponding to site $n=6$. Parameters: $S=0.3$, $B=0.1$, $\theta=4$.}\label{fig04}
	\end{center}
	%\vspace{-5mm}
\end{figure}

To illustrate and further analyze this asymmetry, Fig.~\ref{fig04} shows the TPS corresponding to two sites adjacent to the initially excited site, located on its left- and right-hand sides, respectively. The numerical results are presented for a structure with $M=4$ and $N=16$. Although the sites $n=4$ and $n=6$ are symmetrically positioned with respect to the initially excited site ($n_0=5$), and the corresponding transfer--integrals governing the hopping processes between the pairs of sites $(4,5)$ and $(5,6)$ are identical, the resulting TPSs $p_4(\tau)$ and $p_6(\tau)$ differ, as shown in Fig.~\ref{fig04}. More precisely, the probabilities $p_4(\tau)$ and $p_6(\tau)$ remain practically identical at short times. However, at later times, when the finite size of the MS begins to influence the dynamics, the spectrum corresponding to $p_4(\tau)$ starts to deviate from that of $p_6(\tau)$.

%============================= Fig. 4 =================================
%============================= Fig. 4 =================================

\subsection{Temporal periodicity of the TPS $p_n(\tau)$\\}

%\textcolor{red}{U ovom pod-poglavlju naglasene su dvije uloge $\Omega_k$. 1) same vrijednosti $\Omega_k$ odredjuju residence time; 2) neekvidistantni karakter $\Omega_k$ je odgovoran za nepostojanje revivala! Ovo bi trebali da se ima na umu u zakljucku rada.}\\

\begin{figure}[h]
	\begin{center}
		\includegraphics[width=4.2cm]{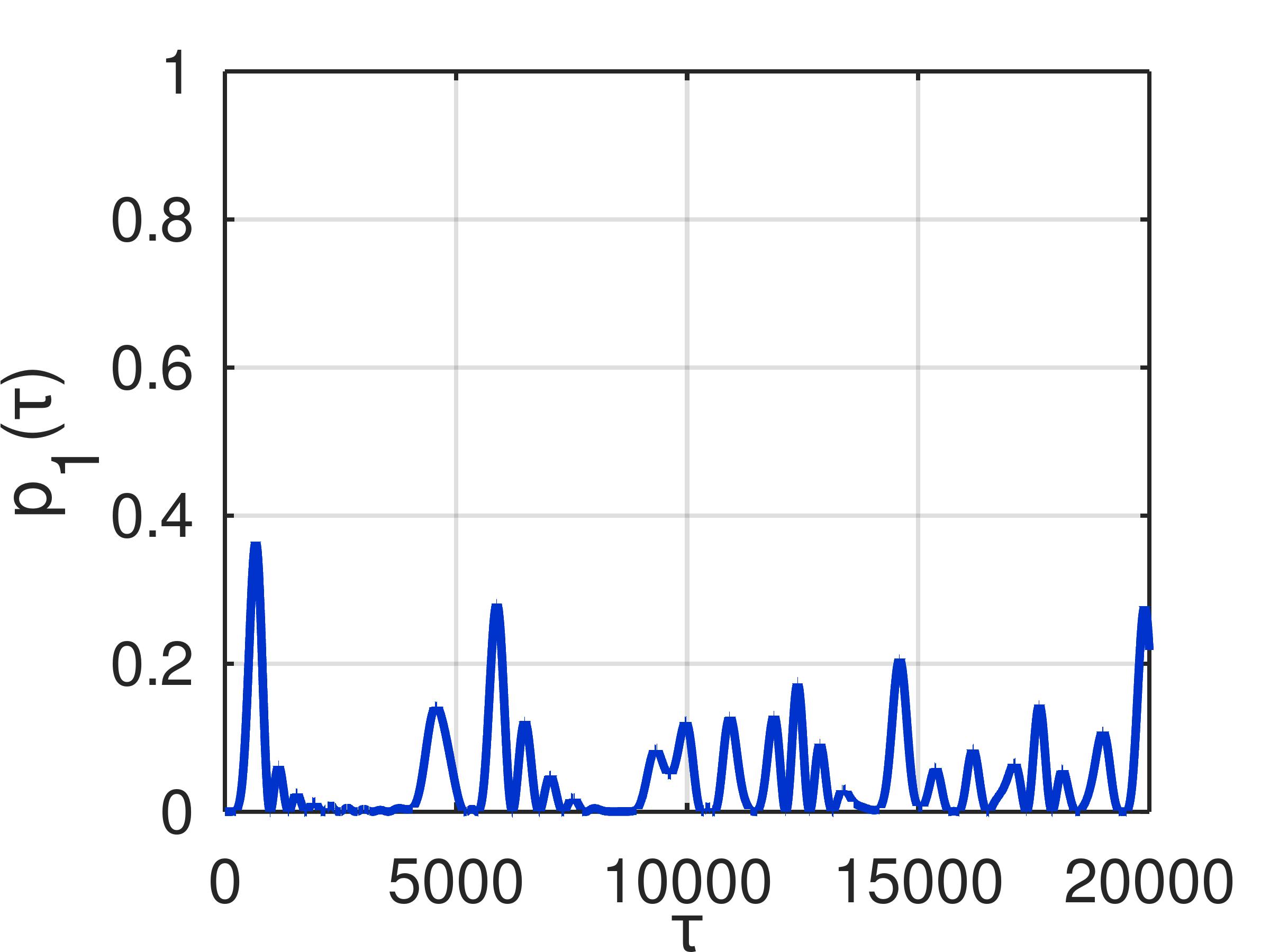}
		\includegraphics[width=4.2cm]{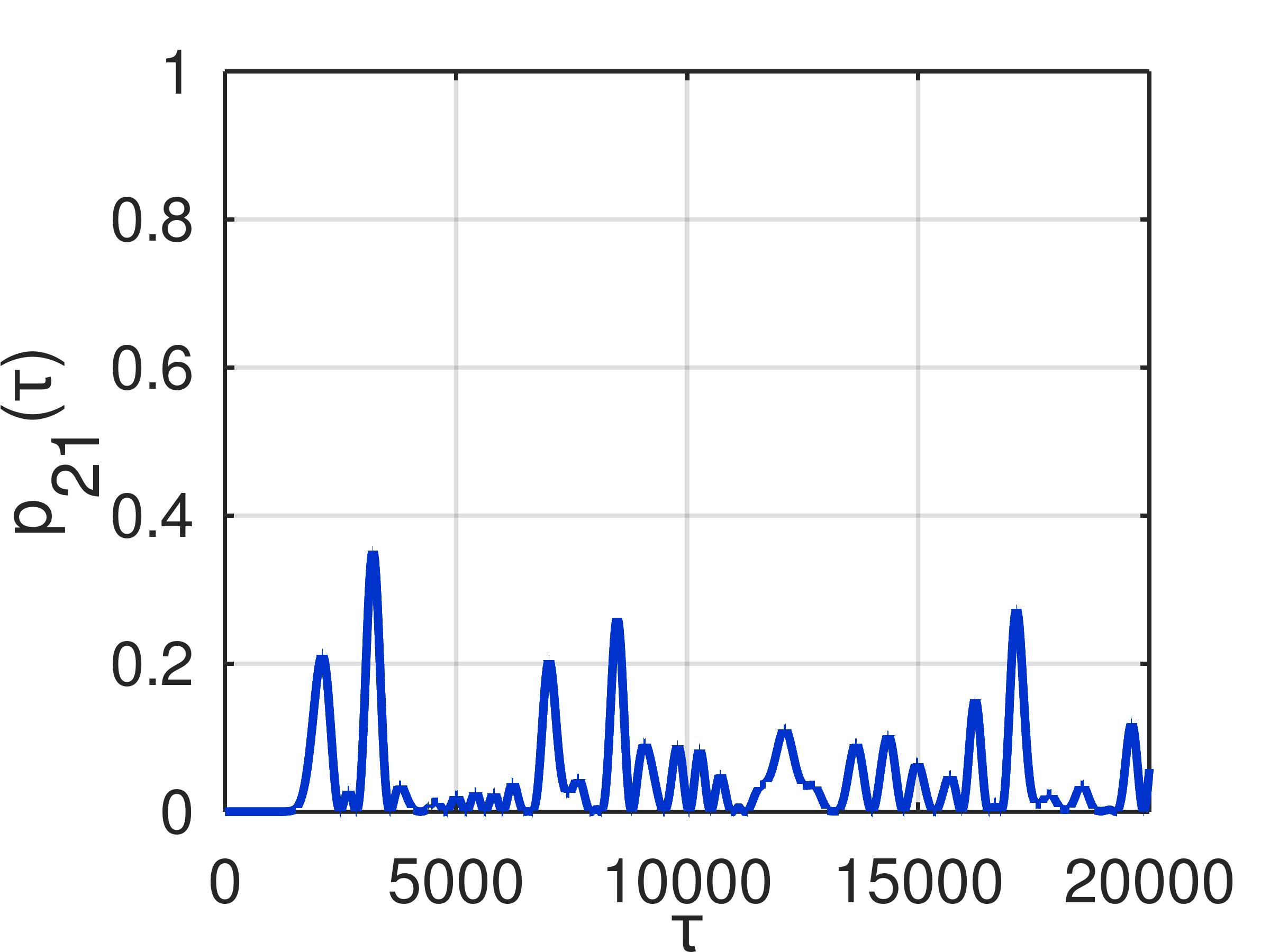}
		\caption{TPS on the edge nodes of the MS, for the configuration $M=4$, $N=16$ ($K=21$) SEs. The node numbering in the graph follows the convention used in Fig.~\ref{fig01}. The initially excited node is $n_0=5$. Left panel: probability distribution for the node $n=1$. Right panel: probability distribution for the node $n=21$. Parameters: $S=0.3$, $B=0.1$, $\theta=4$.}\label{fig05}
	\end{center}
	%\vspace{-5mm}
\end{figure}

From Eqs.~(\ref{VntauN}) and (\ref{pntau}) it follows that the temporal periodicity of the probability distribution $p_n(\tau)$, which determines the spacing between successive maxima/minima, is governed by the frequency spectrum $\Omega_k$ given in Eq.~(\ref{VntauN} b). The minimal period is therefore estimated as $$T_{\min} \sim \frac{2\pi}{B}\mathrm{e}^{S\mathrm{coth}(1/2\theta)},$$ which corresponds to the maximal value $\cos(\phi_k)\sim 1$. This parameter determines the so-called "residence time", i.e., the characteristic time during which the excitation remains localized at a given site and the corresponding structural element of the molecular chain exhibits modified physical properties \cite{CDAIPAdv2026}. For the chosen parameter values $S=0.3$, $B=0.1$, $\theta=4$, one obtains $T_{\min}\sim 700$ (dimensionless units, corresponding to $t=\tau/\omega_0\approx 70\;\text{ps}$), which is consistent with the width of the maxima observed in Figs.~\ref{fig04} and \ref{fig05}. It should be noted that $T_{\min}$ depends exponentially on $S$ and $\theta$, so that relatively small parameter variations can lead to its significant changes. However, $T_{\min}$ only approximately characterizes the separation between adjacent maxima, since $p_n(\tau)$ arises from a coherent superposition of multiple harmonics with different $\Omega_k$, resulting in an interference pattern in which the observed maxima originate from constructive interference of several modes.

The non-equidistant character of the spectrum $\Omega_k$ defined in Eq.~(\ref{VntauN} b) is responsible for the absence of well-defined full revivals in $p_n(\tau)$. Since the phases associated with different modes do not rephase simultaneously, dephasing effects emerge at longer time scales. Nevertheless, due to the discrete nature of the frequencies $\Omega_k$, partial rephasing of subsets of modes can occur at intermediate times, leading to weak and fragmented revival-like structures.

From Fig.~\ref{fig05} one can estimate the characteristic time associated with the propagation of the excitation from the initially excited site toward the left and right boundaries of the molecular segment. This timescale is consistent with the classical propagation time $T_{\mathrm{classical}}$. Beyond this timescale, dephasing effects become significant, and a picture of propagation analogous to a classical particle becomes no longer applicable.

%============================= Fig. 5 =================================
%============================= Fig. 5 =================================

\subsection{Excitation migration through the molecular segment\\}

\begin{figure*}[t]
	\begin{center}
		\subfloat[]{\includegraphics[width=0.32\textwidth]{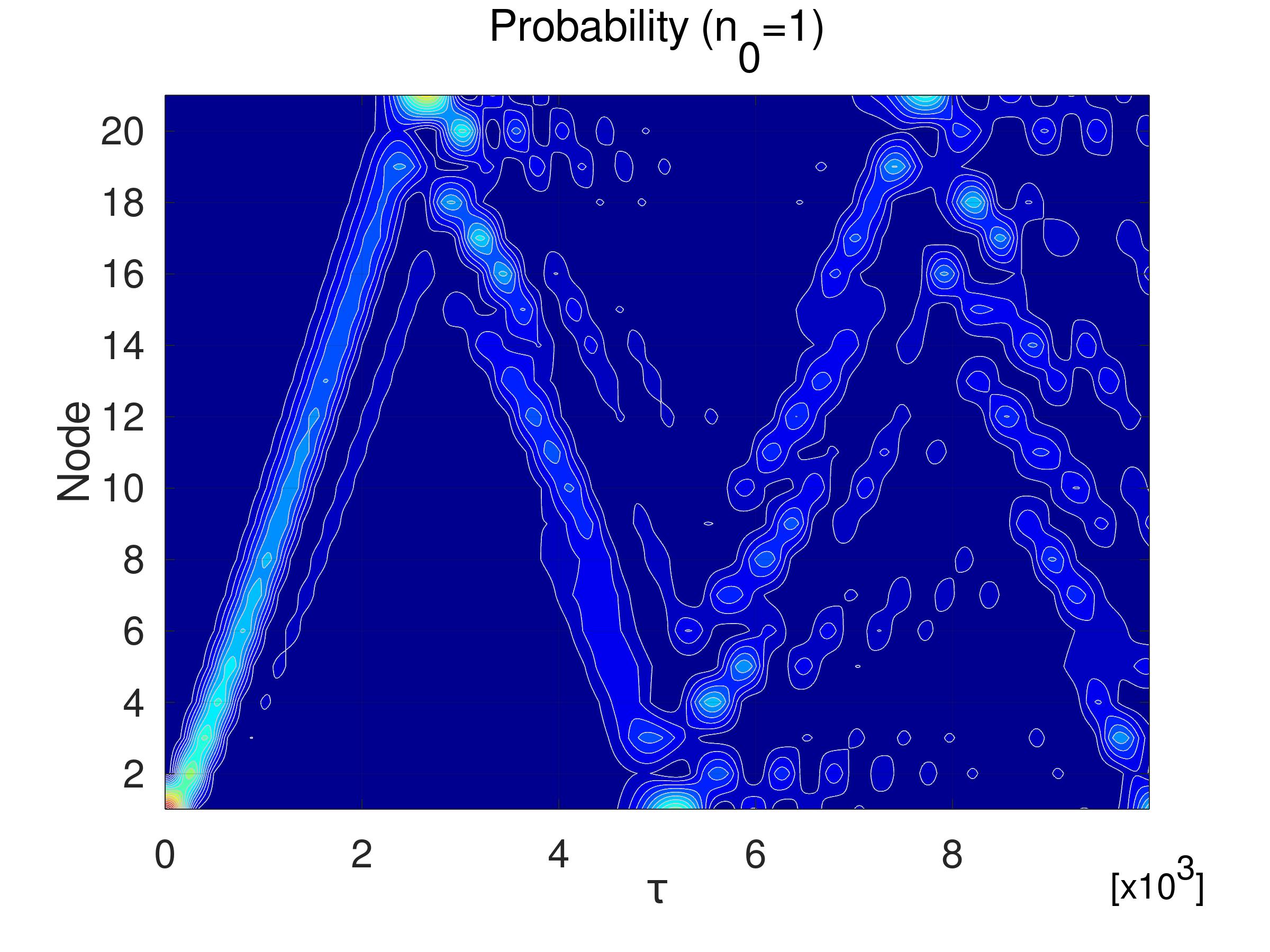}}
		\subfloat[]{\includegraphics[width=0.32\textwidth]{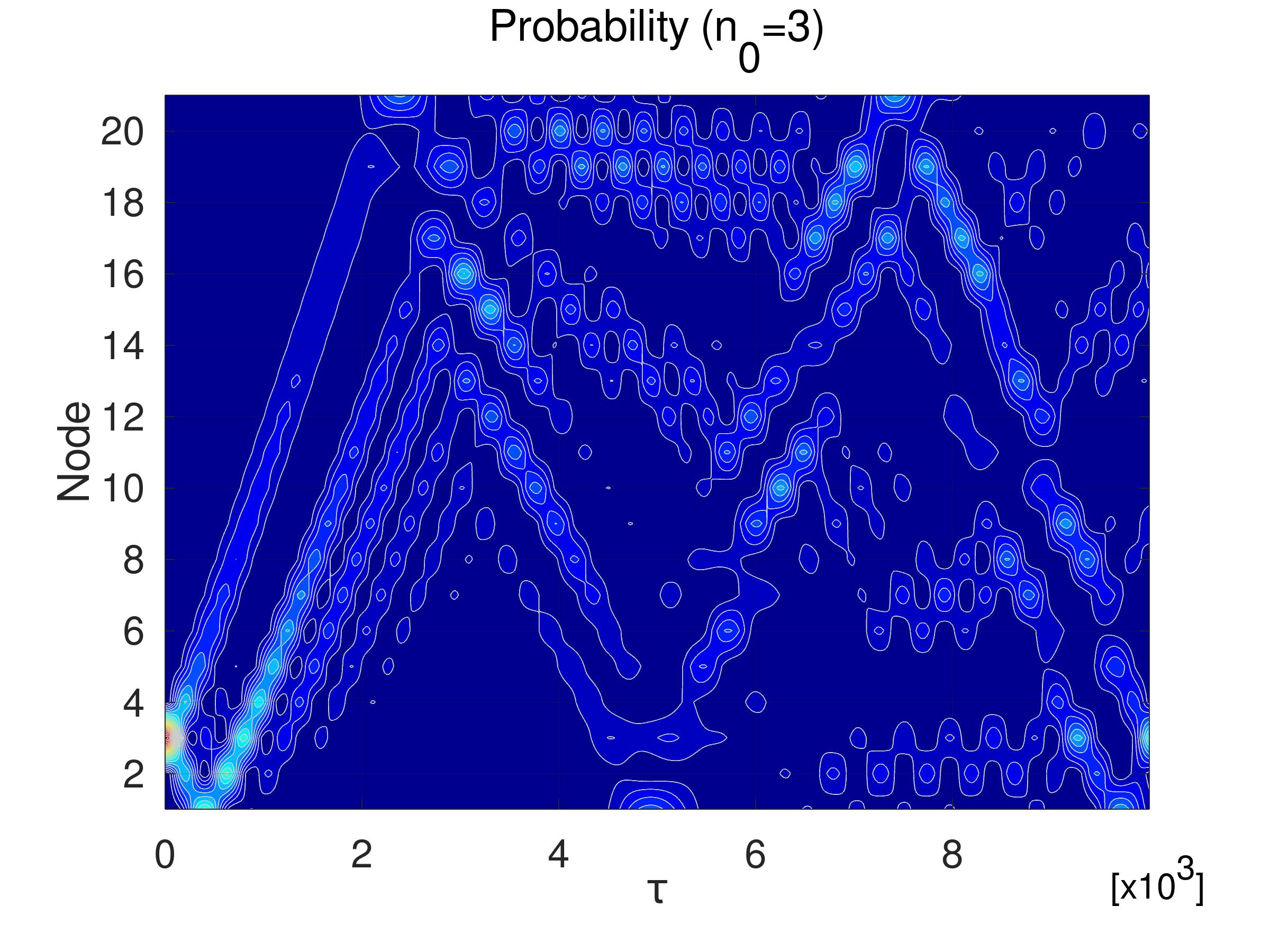}}
		\subfloat[]{\includegraphics[width=0.32\textwidth]{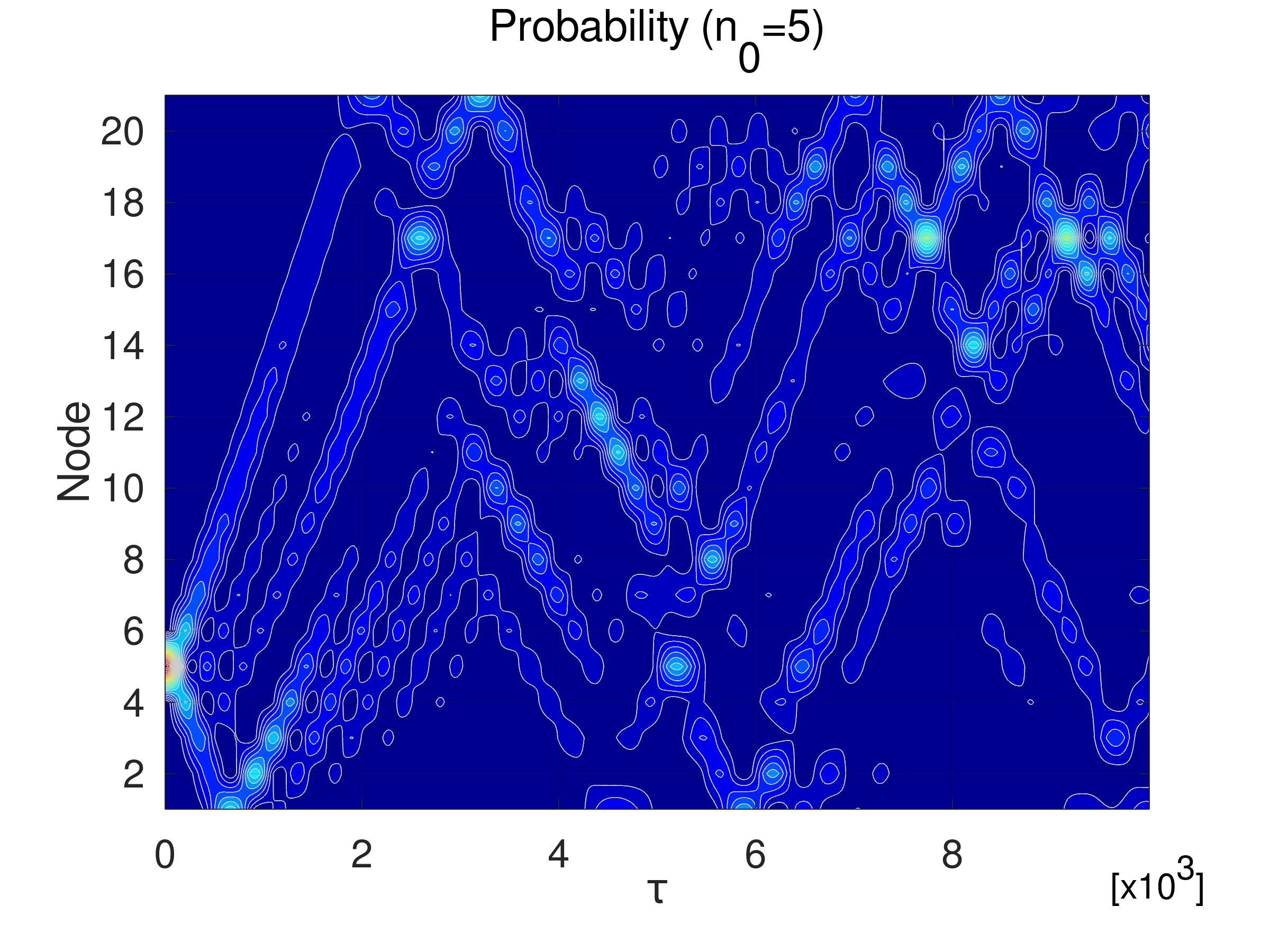}}\\
		\subfloat[]{\includegraphics[width=0.32\textwidth]{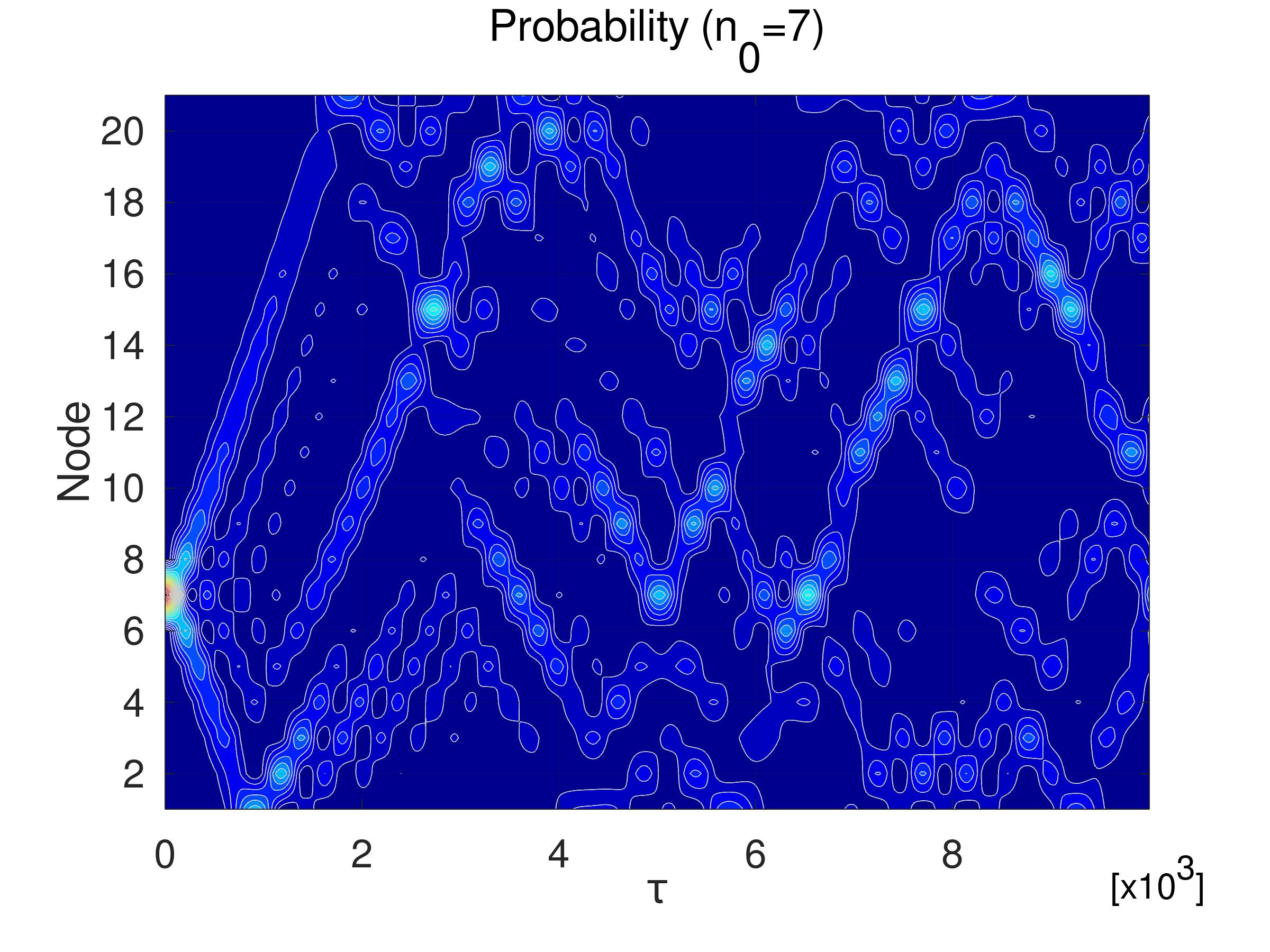}}
		\subfloat[]{\includegraphics[width=0.32\textwidth]{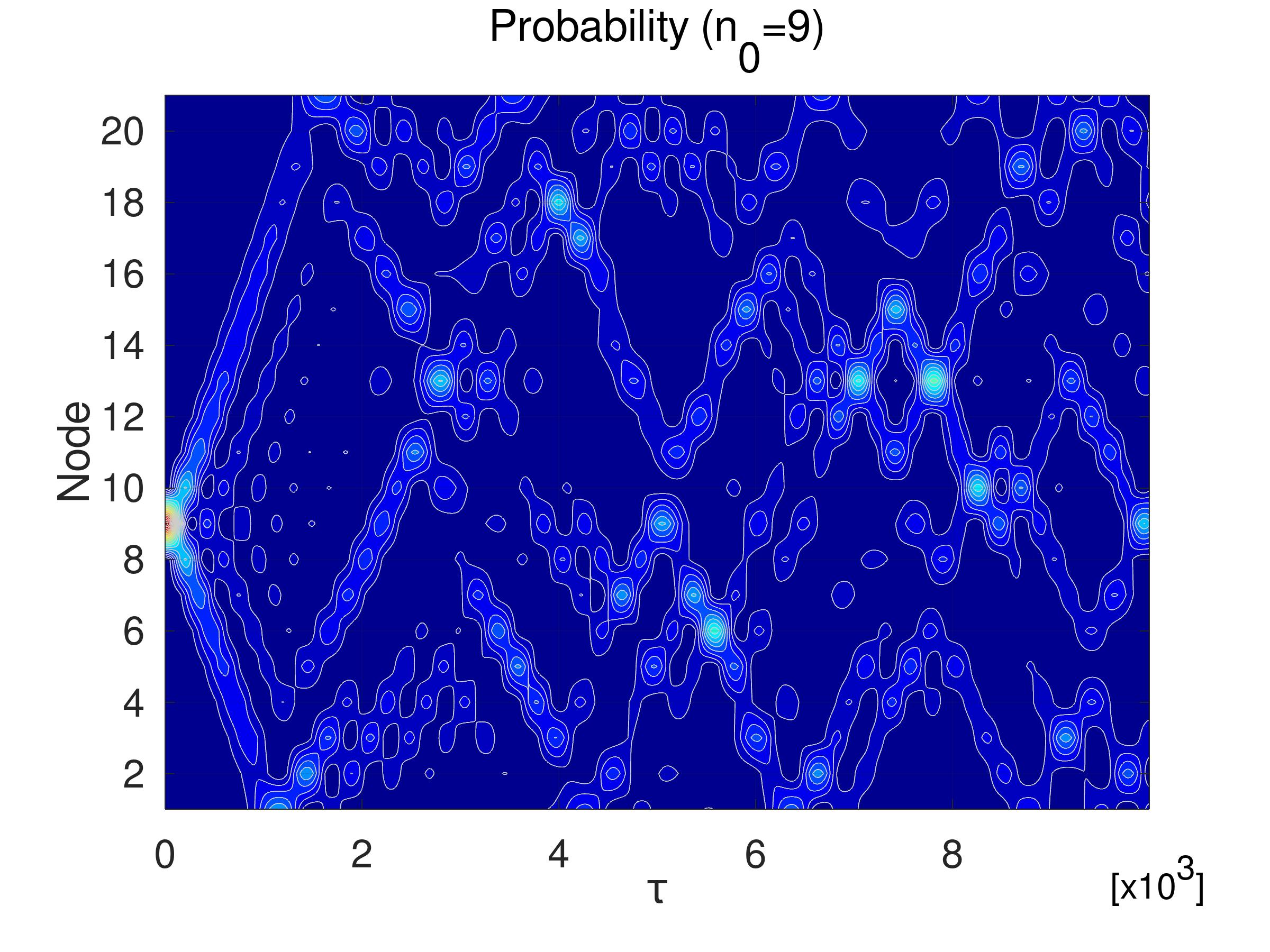}}
		\subfloat[]{\includegraphics[width=0.32\textwidth]{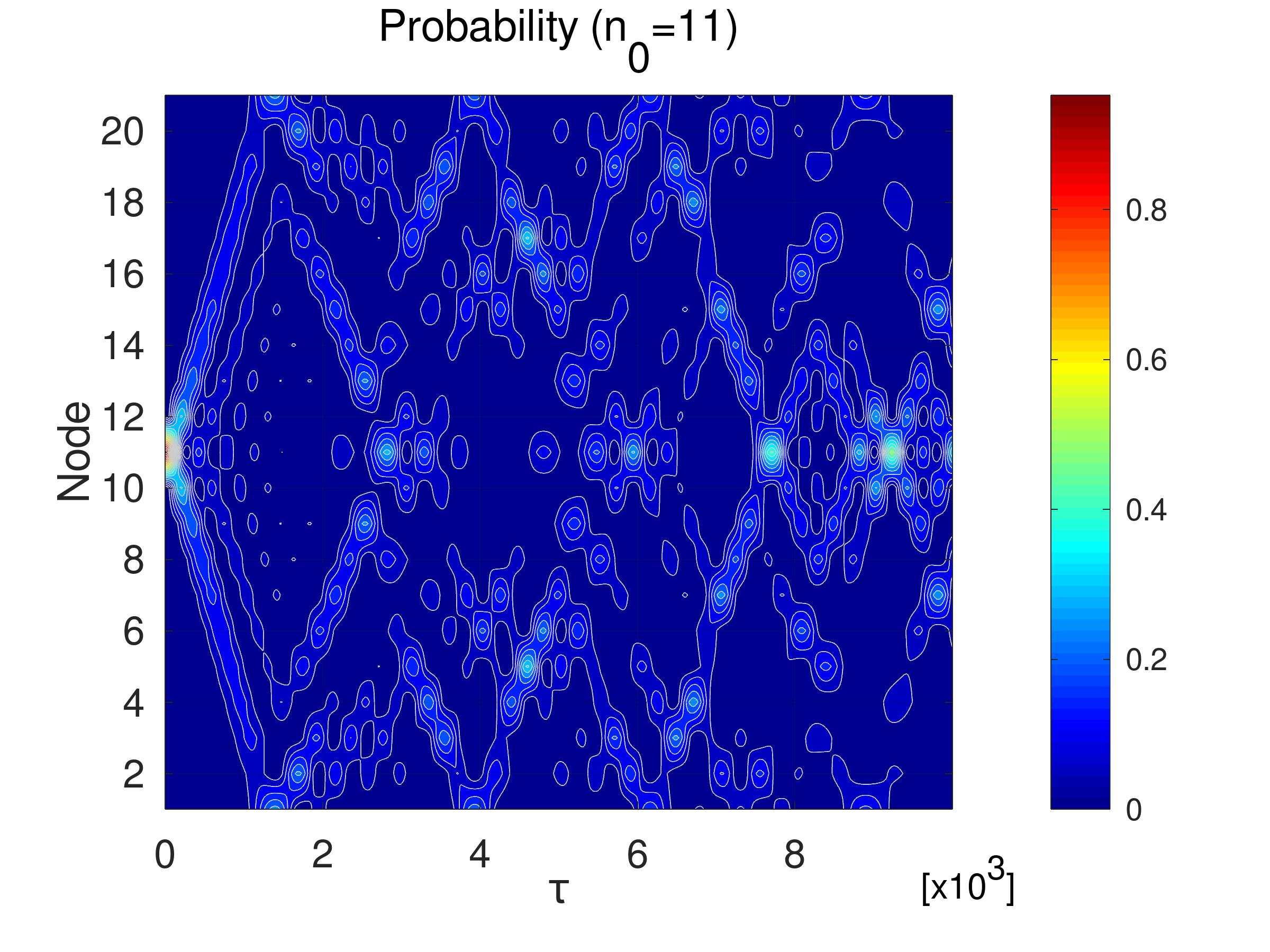}}
		\caption{Excitation probability distribution, in dependence on the position of the initially excited node. Parameters: $S=0.3$, $B=0.1$, $\theta=4$ corresponds to the typical values of the biomolecular chains like alpha--helix protein molecules.}\label{fig06}
	\end{center}
\end{figure*}

In order to analyze the migration of the excitation through the molecular segment, we investigate the excitation probability distribution for different positions of the initially excited site. The simulations were performed for a molecular segment composed of $K=21$ SEs. For clarity and easier interpretation of the obtained results, the site numbering in the figures follows the convention introduced in Fig.~\ref{fig01}. In this way, the position of the initially excited site within the considered structure can be identified directly. The obtained results are presented in panels (a)--(f) of Fig.~\ref{fig06}.

From the presented figures, we observe that the excitation probability distribution $p_n(\tau)$ exhibits an interesting spatiotemporal pattern. The maxima of the probability distribution, arising from the interference of multiple harmonics, initially form a structure that closely resembles the propagation of a classical particle along the molecular segment. When the excitation is initially induced at one edge of the MS (Fig.~\ref{fig06}(a)), the probability maximum gradually propagates toward the opposite edge of the segment (i.e., toward the site numbered as $n=21$). The first pronounced maximum of the distribution $p_{21}(\tau)$ appears at that site after a finite time interval following the excitation injection into the MS (see also the right panel of Fig.~\ref{fig05}). At later times, the probability distribution evolves in the opposite direction, accompanied by a gradual fragmentation of the initially localized probability maximum into multiple secondary maxima. As time evolves further, the initially quasi-localized propagation pattern transforms into a complex interference structure composed of multiple smaller maxima. This behavior reflects the dephasing between different spectral components of the excitation, originating from the non-equidistant character of the underlying frequency spectrum $\Omega_k$. As a consequence, the initially quasi-classical propagation picture is progressively replaced by a fully developed quantum interference regime.

As the initially excited site is shifted toward the center of the molecular segment, the maxima of the distribution $p_n(\tau)$ form a pattern resembling excitation propagation toward both ends of the MS (Fig.~\ref{fig06}, panels (b), (c), and (d)). Similarly to the previously discussed case shown in Fig.~\ref{fig06}(a), the initially well-defined propagation pattern gradually transforms, after successive reflections from the segment boundaries, into an increasingly fragmented probability distribution in which the dominant maxima become accompanied by multiple secondary maxima of lower intensity. The fragmentation of the excitation probability distribution therefore becomes progressively more pronounced. Nevertheless, the obtained results indicate that the more pronounced probability maxima tend to occur at sites located farther from the initially excited site than at those in its immediate vicinity. In particular, the maxima of the probability distribution at the edge of the MS that is located closer to the initially excited site remain slightly lower than those corresponding to the opposite edge.

As the initially excited site approaches the center of the chain, the probability distribution becomes increasingly fragmented and interference-dominated (Fig.~\ref{fig06}, panels (e) and (f)). At the same time, the sites associated with the most pronounced probability maxima gradually shift from the segment edges toward the central region of the chain. When the excitation is initially induced exactly at the center of the molecular segment, the probability distribution becomes symmetric with respect to this site, and the highest-probability sites are arranged symmetrically around it (Fig.~\ref{fig06}(f)).

%============================= Fig. 6 =================================
%============================= Fig. 6 =================================

The results presented in Fig.~\ref{fig06}, similarly to the results presented in Fig.~\ref{fig04}, indicate that the position of the initially excited site within the MS can induce an asymmetric excitation probability distribution with respect to the initially excited site. However, unlike the asymmetry discussed in connection with Fig.~\ref{fig04}, which has a local character and is primarily associated with the immediate vicinity of the initially excited site, the asymmetry observed here is of a global nature, involving the two subsegments defined by the position of the initially excited site.

\begin{figure}[h]
	\begin{center}
		\includegraphics[width=6cm]{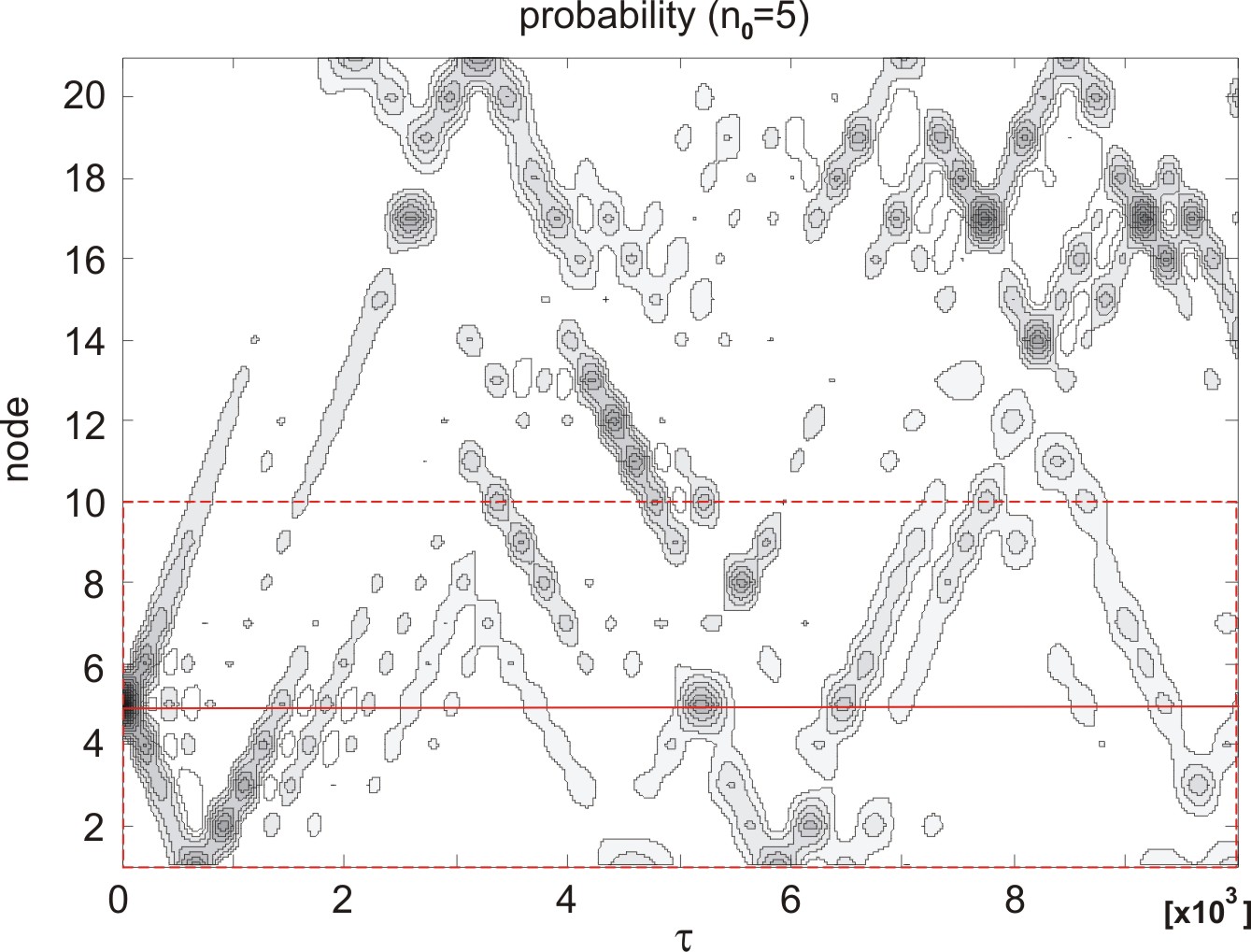}
		\caption{Time distribution of the excitation probability over the nodes of the MS. Parameters: $S=0.3$, $B=0.1$, $\theta=4$. Excitation was induced at the fifth SE from the left edge of the MS. Rectangled area is divided into the two symmetric regions, corresponding to fifth SE above the initially excited node, ant the fifith SE below the initially excited node.}\label{fig07}
	\end{center}
	%\vspace{-5mm}
\end{figure}

Indeed, if we consider the excitation probability distribution along the MS when the excitation is initially injected into a site located inside the segment but away from its center (for example, $n_0=5$ as shown in Fig.~\ref{fig07}), we observe that the probability of finding the excitation is generally higher on the longer subsegment ("above" the solid line in Fig.~\ref{fig07}) than on the shorter one. More precisely, the longer subsegment exhibits a larger number of pronounced probability maxima, whose intensities also tend to increase slightly with time. This tendency remains visible even when the comparison is restricted to pairs of sites symmetrically positioned with respect to the initially excited site and belonging to the longer and shorter subsegments, respectively (see the probability distributions shown within the rectangle in Fig.~\ref{fig07}).

Similarly to the asymmetry discussed in connection with Eq.~(\ref{VntauN}), this behavior bears some resemblance to quantum ratcheting effects. Although it does not imply the existence of a net excitation current through the molecular segment, it nevertheless suggests the emergence of an effective preferred direction of excitation migration.

%============================= Fig. 7 =================================
%============================= Fig. 7 =================================

\section{Conclusion}

The temporal distributions of the probability of finding an excitation at the nodes of the molecular segment studied here arise from coherent quantum effects within the considered structure. Mathematically, these distributions are the result of a superposition of discrete modes corresponding to the spectrum $\Omega_k$, as defined by Eq.~(\ref{VntauN} b). Consequently, the "migration of the excitation" should be interpreted as a manifestation of the time-dependent interference of these modes, rather than as the propagation of a "classical" particle. We now highlight the following key findings of our study:

\begin{enumerate}
	\item The formal analogy with CTQW models established here could offer deeper insight into the processes of coherent excitation migration through the BmC. The influence of temperature enters the model through the polaron effect; it does not lead to a loss of coherence of the quantum excitation state, but rather merely modifies the transfer integral. This stands in contrast to existing temperature-dependent CTQW approaches, where temperature typically induces decoherence \cite{MosheiniJCP}. Of course, this is valid only for systems where the residual interaction between the polaron and the renormalized phonons is weak. On the other hand, the temperature dependence of the transfer integral predicted by our model may bring CTQW descriptions closer to reality, thereby bridging the gap between idealized quantum-walk theory and experimentally relevant molecular environments.
	
	\item A notable observation is the asymmetry in the probability distribution of finding the excitation at nodes symmetrically positioned with respect to the initially excited site. This asymmetry arises solely from the asymmetric placement of the initially excited node within a finite segment with reflecting boundaries, demonstrating that global geometric asymmetry can induce nontrivial directional features even in locally symmetric quantum transport systems. From a physical standpoint, this is a purely quantum effect, originating from the interplay between the finite geometry of the structure--in particular, the position of the initially excited site--and the boundary conditions. The symmetry breaking is not in the Hamiltonian itself, but in the initial condition relative to the boundaries, leading to different interference patterns at equidistant sites on the left and right sides, which resemble a boundary-induced ratchet-like effect. However, unlike conventional ratchet systems that generate a net directed current, here the ratchet effect manifests solely as a left--right asymmetry in the quantum interference pattern, without any external driving or dissipative mechanism that would produce a classical directed current. Importantly, this local asymmetry is not confined to the immediate vicinity of the initially excited site; through the coherent coupling mediated by the transfer integral, it propagates throughout the chain, giving rise to a global manifestation of the symmetry breaking as well. The detailed analysis of these local and global manifestations, presented in the Sec.~\ref{Results}, reveals that the longer sub--segment of the MS exhibits a larger number of pronounced probability maxima, suggesting an effective preferred direction of excitation migration even in the absence of a net directed current. This finding highlights the importance of boundary conditions and initial-state geometry in controlling quantum transport in finite molecular systems, and may open new avenues for designing molecular architectures with tailored excitation-transfer properties.
	
	\item The complex interference pattern and the absence of well-defined revivals stem from the non-equidistant spectrum of mode frequencies $\Omega_k$. As a result, the constituent modes associated with different $\Omega_k$ progressively dephase and never fully realign to reconstruct a single dominant probability maximum. Instead, only partial constructive interference occurs, leading to the observed fragmented probability distribution.
	
	\item Consequently, the physical implications for the excitation migration are as follows. Upon initial excitation, the excitation delocalizes onto neighbouring sites on both sides and propagates through the structure. During the early stage, the temporal probability distribution resembles that of a free particle, with a well-defined propagation front. This picture holds until the excitation reaches the boundaries of the segment at times $T^{(L)}_{\text{classical}}$ and $T^{(R)}_{\text{classical}}$ for the left and right ends, respectively. After reflection from the boundaries, the temporal probability distribution at a given site becomes significantly more complex: instead of a single pronounced maximum, it exhibits one dominant maximum accompanied by several secondary maxima of lower intensity. In other words, the initially coherent sum in Eq.~(\ref{VntauN}) undergoes progressive dephasing, leading to a fragmented probability distribution. Consequently, the simple picture of the excitation propagating as a classical-like particle is lost, and the dynamics enter a regime characterized by intricate interference patterns.
\end{enumerate}

Presented findings demonstrate that the coherent excitation dynamics in finite molecular segments is governed by a rich interplay between quantum interference, boundary conditions, and the initial excitation geometry. Our analytical approach, based on the Lang-Firsov transformation and mean-field averaging, not only provides a rigorous framework for understanding these phenomena but also establishes a direct link between molecular excitation transport and the well-established theory of continuous-time quantum walks. We believe that the temperature-dependent transfer integral derived here may serve as a valuable tool for bridging the gap between idealized quantum-walk models and realistic molecular systems, paving the way for future experimental investigations of coherent transport in biomolecular and organic materials.

%\onecolumn
%	\begin{figure}[h]
%		\begin{center}
%			\includegraphics[width=5.6cm]{K21p1CB}
%			\includegraphics[width=5.6cm]{K21p3CB}
%			\includegraphics[width=5.6cm]{K21p5CB}\\
%			\vskip 0.4cm
%			\includegraphics[width=5.6cm]{K21p7CB}
%			\includegraphics[width=5.6cm]{K21p9CB}
%			\includegraphics[width=5.6cm]{K21p11CB}
%			%\vspace{-3mm}
%			\caption{Time distribution of the excitation probability on the nodes of the MC. Parameters: $S=0.3$, $B=0.1$, $\theta=4$. Excitation was induced at the second SE from the left.}\label{fig06}
%		\end{center}
%			%\vspace{-5mm}
%	\end{figure}
%\twocolumn

\section{Acknowledgment}

This work was supported by the Ministry of Science, Technological Development, and Innovation of the Republic of Serbia through the Project contract No 451-03-33/2026-03/200017.

\bibliographystyle{apsrev4-2}
\bibliography{references}

@book{Voet,
  author    = {Donald Voet and Judith G. Voet},
  title     = {Biochemistry},
  edition   = {3},
  publisher = {Wiley},
  address   = {New York},
  year      = {2004}
}

@book{Lehninger,
  author    = {David L. Nelson and Michael M. Cox},
  title     = {Lehninger Principles of Biochemistry},
  publisher = {W. H. Freeman},
  address   = {New York},
  year      = {2013}
}

@book{Dauxois,
  author    = {Thierry Dauxois and Michel Peyrard},
  title     = {Physics of Solitons},
  publisher = {Cambridge University Press},
  address   = {Cambridge},
  year      = {2006}
}

@book{Frohlich,
  editor    = {H. Fr{\"o}hlich and F. Kremer},
  title     = {Coherent Excitations in Biological Systems},
  publisher = {Springer},
  address   = {Berlin},
  year      = {1983}
}

@article{LambertNP2013,
  author  = {N. Lambert and Y. N. Chen and Y. C. Cheng and C. M. Li and G. Y. Chen and F. Nori},
  title   = {Quantum biology},
  journal = {Nature Physics},
  volume  = {9},
  pages   = {10--18},
  year    = {2013},
  doi     = {10.1038/nphys2474}
}

@article{ChenACIE,
  author  = {Xi Chen and Xue Zhang and Xiao Xiao and Zhijia Wang and Jianzhang Zhao},
  title   = {Recent Developments on Understanding Charge Transfer in Molecular Electron Donor--Acceptor Systems},
  journal = {Angewandte Chemie International Edition},
  volume  = {62},
  pages   = {e202216010},
  year    = {2023},
  doi     = {10.1002/anie.202216010}
}

@article{CruzeiroLTP,
  author  = {L. Cruzeiro},
  title   = {Knowns and unknowns in the Davydov model for energy transfer in proteins},
  journal = {Low Temperature Physics},
  volume  = {48},
  pages   = {973},
  year    = {2022},
  doi     = {10.1063/10.0015107}
}

@article{AlvarezFQST2024,
  author  = {P. H. Alvarez and L. Gerhards and I. A. Solov'yov and M. C. de Oliveira},
  title   = {Quantum phenomena in biological systems},
  journal = {Frontiers in Quantum Science and Technology},
  volume  = {3},
  pages   = {1466906},
  year    = {2024},
  doi     = {10.3389/frqst.2024.1466906}
}

@article{CDAIPAdv2026,
  author  = {D. Chevizovich and S. Galovic and V. Matic and Z. Ivic and Z. Przulj},
  title   = {Influence of induced excitation on the functionality of finite-length molecular chain},
  journal = {AIP Advances},
  volume  = {16},
  pages   = {025306},
  year    = {2026},
  doi     = {10.1063/5.0309310}
}

@incollection{ZdCeND,
  author    = {D. Chevizovich},
  editor    = {S. Zdravkovic and D. Chevizovich},
  title     = {Vibron Self-trapping in Quasi-One-Dimensional Biomolecules: Non-adiabatic Polaron Approach},
  booktitle = {Nonlinear Dynamics of Nanobiophysics},
  publisher = {Springer Nature},
  address   = {Singapore},
  year      = {2022},
  doi       = {10.1007/978-981-19-5323-1_8}
}

@article{HolsteinAP1,
  author  = {T. Holstein},
  title   = {Studies of polaron motion: Part I. The molecular-crystal model},
  journal = {Annals of Physics},
  volume  = {8},
  pages   = {325--342},
  year    = {1959},
  doi     = {10.1016/0003-4916(59)90002-8}
}

@article{HolsteinAP2,
  author  = {T. Holstein},
  title   = {Studies of polaron motion: Part II. The ``small'' polaron},
  journal = {Annals of Physics},
  volume  = {8},
  pages   = {343--389},
  year    = {1959},
  doi     = {10.1016/0003-4916(59)90003-X}
}

@article{LF,
  author  = {I. G. Lang and Yu. A. Firsov},
  title   = {Kinetic Theory of Semiconductors with Low Mobility},
  journal = {Soviet Physics JETP},
  volume  = {16},
  pages   = {1301--1312},
  year    = {1963}
}

@article{AK,
  author  = {D. M. Alexander and J. A. Krumhansl},
  title   = {Localized excitations in hydrogen-bonded molecular crystals},
  journal = {Physical Review B},
  volume  = {33},
  pages   = {7172--7184},
  year    = {1986},
  doi     = {10.1103/PhysRevB.33.7172}
}

@article{PouthierJCP132,
  author  = {V. Pouthier},
  title   = {Vibron phonon in a lattice of H-bonded peptide units: A criterion to discriminate between the weak and the strong coupling limit},
  journal = {Journal of Chemical Physics},
  volume  = {132},
  pages   = {035106},
  year    = {2010},
  doi     = {10.1063/1.3297947}
}

@article{PouthierPRL,
  author  = {J. Edler and R. Pfister and V. Pouthier and C. Falvo and P. Hamm},
  title   = {Direct Observation of Self-Trapped Vibrational States in $\alpha$-Helices},
  journal = {Physical Review Letters},
  volume  = {93},
  pages   = {106405},
  year    = {2004},
  doi     = {10.1103/PhysRevLett.93.106405}
}

@article{CareriPRL51,
  author  = {G. Careri and U. Buontempo and F. Carta and E. Gratton and A. C. Scott},
  title   = {Infrared Absorption in Acetanilide by Solitons},
  journal = {Physical Review Letters},
  volume  = {51},
  pages   = {304},
  year    = {1983},
  doi     = {10.1103/PhysRevLett.51.304}
}

@article{BlanchetPRL54,
  author  = {G. B. Blanchet and C. R. Fincher Jr.},
  title   = {Defects in a Nonlinear Pseudo One-Dimensional Solid},
  journal = {Physical Review Letters},
  volume  = {54},
  pages   = {1310},
  year    = {1985},
  doi     = {10.1103/PhysRevLett.54.1310}
}

@article{HammTsironisEPJST147,
  author  = {P. Hamm and G. P. Tsironis},
  title   = {Semiclassical and quantum polarons in crystalline acetanilide},
  journal = {European Physical Journal Special Topics},
  volume  = {147},
  pages   = {303--331},
  year    = {2007},
  doi     = {10.1140/epjst/e2007-00215-7}
}

@article{HammTsironis,
  author  = {P. Hamm and G. P. Tsironis},
  title   = {Barrier crossing to the small Holstein polaron regime},
  journal = {Physical Review B},
  volume  = {78},
  pages   = {092301},
  year    = {2008},
  doi     = {10.1103/PhysRevB.78.092301}
}

@article{PouthierPRE2008,
  author  = {V. Pouthier},
  title   = {Amide-I lifetime-limited vibrational energy flow in a one-dimensional lattice of hydrogen-bonded peptide units},
  journal = {Physical Review E},
  volume  = {78},
  pages   = {061909},
  year    = {2008},
  doi     = {10.1103/PhysRevE.78.061909}
}

@article{CastroNetoCaldeira,
  author  = {A. H. Castro Neto and A. O. Caldeira},
  title   = {Alternative approach to the dynamics of polarons in one dimension},
  journal = {Physical Review B},
  volume  = {46},
  pages   = {8858},
  year    = {1992},
  doi     = {10.1103/PhysRevB.46.8858}
}

@article{DevreeseRPP2009,
  author  = {J. T. Devreese and A. S. Alexandrov},
  title   = {Fr{\"o}hlich polaron and bipolaron: recent developments},
  journal = {Reports on Progress in Physics},
  volume  = {72},
  pages   = {066501},
  year    = {2009},
  doi     = {10.1088/0034-4885/72/6/066501}
}

@article{KalosakasPRB,
  author  = {G. Kalosakas and S. Aubry and G. P. Tsironis},
  title   = {Polaron solutions and normal-mode analysis in the semiclassical Holstein model},
  journal = {Physical Review B},
  volume  = {58},
  pages   = {3094},
  year    = {1998},
  doi     = {10.1103/PhysRevB.58.3094}
}

@article{Barthes1989,
  author  = {M. Barthes},
  title   = {Optical anomalies in acetanilide: Davydov solitons, localised modes, or Fermi resonance?},
  journal = {Journal of Molecular Liquids},
  volume  = {41},
  pages   = {143},
  year    = {1989},
  doi     = {10.1016/0167-7322(89)80075-3}
}

@article{Edler2002,
  author  = {J. Edler and P. Hamm and A. C. Scott},
  title   = {Femtosecond Study of Self-Trapped Vibrational Excitons in Crystalline Acetanilide},
  journal = {Physical Review Letters},
  volume  = {88},
  pages   = {067403},
  year    = {2002},
  doi     = {10.1103/PhysRevLett.88.067403}
}

@article{YarkonyJCP,
  author  = {D. Yarkony and R. Silbey},
  title   = {Comments on exciton-phonon coupling: Temperature dependence},
  journal = {Journal of Chemical Physics},
  volume  = {65},
  pages   = {1042},
  year    = {1976},
  doi     = {10.1063/1.433182}
}

@article{HennigPRB,
  author  = {D. Hennig},
  title   = {Energy transport in $\alpha$-helical protein models: One-strand versus three-strand systems},
  journal = {Physical Review B},
  volume  = {65},
  pages   = {174302},
  year    = {2002},
  doi     = {10.1103/PhysRevB.65.174302}
}

@incollection{Rashba,
  author    = {E. I. Rashba},
  editor    = {E. I. Rashba and M. Sturge},
  title     = {Excitons},
  publisher = {North-Holland},
  address   = {Amsterdam},
  year      = {1982}
}

@book{Mahan,
  author    = {G. D. Mahan},
  title     = {Many-Particle Physics},
  publisher = {Plenum Press},
  address   = {New York},
  year      = {1986}
}

@article{Nevskaya,
  author  = {N. A. Nevskaya and Yu. N. Chirgadze},
  title   = {Infrared spectra and resonance interactions of amide-I and II vibrations of $\alpha$-helix},
  journal = {Biopolymers},
  volume  = {15},
  pages   = {637--648},
  year    = {1976},
  doi     = {10.1002/bip.1976.360150404}
}

@article{FalvoPouthier,
  author  = {C. Falvo and V. Pouthier},
  title   = {Vibron-polaron in $\alpha$-helices. I. Single-vibron states},
  journal = {Journal of Chemical Physics},
  volume  = {123},
  pages   = {184709},
  year    = {2005},
  doi     = {10.1063/1.2101569}
}

@article{HammEdlerPRB73,
  author  = {P. Hamm and J. Edler},
  title   = {Quantum vibrational polarons: Crystalline acetanilide revisited},
  journal = {Physical Review B},
  volume  = {73},
  pages   = {094302},
  year    = {2006},
  doi     = {10.1103/PhysRevB.73.094302}
}

@article{KalosakasPRE,
  author  = {G. Kalosakas},
  title   = {Charge transport in DNA: Dependence of diffusion coefficient on temperature and electron-phonon coupling constant},
  journal = {Physical Review E},
  volume  = {84},
  pages   = {051905},
  year    = {2011},
  doi     = {10.1103/PhysRevE.84.051905}
}

@article{Kempe2003,
  author  = {J. Kempe},
  title   = {Quantum random walks: An introductory overview},
  journal = {Contemporary Physics},
  volume  = {44},
  pages   = {307--327},
  year    = {2003},
  doi     = {10.1080/00107151031000110776}
}

@article{MB2011,
  author  = {O. M{\"u}lken and A. Blumen},
  title   = {Continuous-time quantum walks: Models for coherent transport on complex networks},
  journal = {Physics Reports},
  volume  = {502},
  pages   = {37--87},
  year    = {2011},
  doi     = {10.1016/j.physrep.2011.01.002}
}

@misc{FussArXiv2007,
  author       = {I. Fuss and L. White and P. Sherman and S. Naguleswaran},
  title        = {An analytic solution for one-dimensional quantum walks},
  year         = {2007},
  eprint       = {0705.0077},
  archivePrefix= {arXiv},
  primaryClass = {quant-ph},
  doi          = {10.48550/arXiv.0705.0077}
}

@misc{AperarXiv2024,
  author       = {S. Apers and L. Miclo},
  title        = {Quantum walks, the discrete wave equation and Chebyshev polynomials},
  year         = {2024},
  eprint       = {2402.07809},
  archivePrefix= {arXiv},
  doi          = {10.48550/arXiv.2402.07809}
}

@article{PhysRepReiman2002,
  author  = {P. Reimann},
  title   = {Brownian motors: noisy transport far from equilibrium},
  journal = {Physics Reports},
  volume  = {361},
  pages   = {57--265},
  year    = {2002},
  doi     = {10.1016/S0370-1573(01)00081-3}
}

@article{Kozyrev2023,
  author  = {S. V. Kozyrev and A. N. Pechen},
  title   = {Amplification of quantum transfer and quantum ratchet},
  journal = {Physica Scripta},
  volume  = {98},
  pages   = {125122},
  year    = {2023},
  doi     = {10.1088/1402-4896/ad0c3d}
}

@article{LauRSC2017,
  author  = {B. Lau and O. Kedem and J. Schwabacher and D. Kwasnieski and E. A. Weiss},
  title   = {An introduction to ratchets in chemistry and biology},
  journal = {Materials Horizons},
  volume  = {4},
  pages   = {310--318},
  year    = {2017},
  doi     = {10.1039/C7MH00062F}
}

@article{MosheiniJCP,
	author = {Mohseni, Masoud and Rebentrost, Patrick and Lloyd, Seth and Aspuru-Guzik, Alán},
	title = {Environment-assisted quantum walks in photosynthetic energy transfer},
	journal = {The Journal of Chemical Physics},
	volume = {129},
	number = {17},
	pages = {174106},
	year = {2008},
	month = {11},
	issn = {0021-9606},
	doi = {10.1063/1.3002335}
}

\end{document}